\documentstyle[seceq,preprint]{jpsj}

\def\gsim{\,$\raise0.3ex\hbox{$>$}\llap{\lower0.8ex\hbox{$\sim$}}$\,}
\def\lsim{\,$\raise0.3ex\hbox{$<$}\llap{\lower0.8ex\hbox{$\sim$}}$\,}
\def\lrarrow{\,$\raise0.4ex\hbox{$\rightarrow$}\llap{\lower0.4ex\hbox{$\leftarrow$}}$\,}

\def\vecS{{\vec S}}
\def\vecs{{\vec s}}
\def\veccalS{{\vec{\cal S}}}
\def\vecSigma{{\vec \Sigma}}
\def\vecsigma{{\vec \sigma}}
\def\vectau{{\vec \tau}}

\def\runtitle{Ground-State and Thermodynamic Properties of the
Quantum Mixed Spin-1/2$\,$-1/2$\,$-1$\,$-1 Chain}
\def\runauthor{Takashi {\sc Tonegawa}, Toshiya {\sc Hikihara}, Makoto
{\sc Kaburagi}, Tomotoshi {\sc Nishino}, Seiji {\sc Miyashita} and
Hans-J{\"u}rgen {\sc Mikeska}}

\title
{
Ground-State and Thermodynamic Properties of\\
the Quantum Mixed Spin-1/2$\,$-1/2$\,$-1$\,$-1 Chain
}

\author
{
Takashi {\sc Tonegawa}\footnote{tonegawa@kobe-u.ac.jp}, Toshiya
{\sc Hikihara}$^{1,}$\footnote{hikihara@phys560.phys.kobe-u.ac.jp}, Makoto
{\sc Kaburagi}$^{2}$, Tomotoshi {\sc Nishino},\\
Seiji {\sc Miyashita}$^{3}$ and Hans-J{\"u}rgen {\sc Mikeska}$^{4}$
}

\inst
{
Department of Physics, Faculty of Science, Kobe University, Rokkodai,
Kobe 657\\
$^1$Division of Information and Media Science, Graduate School of Science and
Technology, Kobe University, Rokkodai, Kobe 657\\
$^2$Department of Informatics, Faculty of Cross-Cultural Studies,
Kobe University, Tsurukabuto, Kobe 657\\
$^3$Department of Earth and Space Science, Graduate School of Science, Osaka
University, Toyonaka, Osaka 560\\
$^4$Institut f{\"u}r Theoretische Physik, Universit{\"a}t Hannover, 30167
Hannover, Germany
}

\recdate
{October 6, 1997}

\abst
{
We investigate both analytically and numerically the ground-state and
thermodynamic properties of the quantum mixed spin-1/2$\,$-1/2$\,$-1$\,$-1
chain described by the Hamiltonian \hbox{${\cal H}\!=\!\sum_{\ell=1}^{N/4}\bigl(J_1\,\vecs_{4\ell-3}\cdot\vecs_{4\ell-2}\!+\!J_2\,\vecs_{4\ell-2}\cdot\vecS_{4\ell-1}\!+\!J_3\,\vecS_{4\ell-1}\cdot\vecS_{4\ell}\!+\!J_2\,\vecS_{4\ell}\cdot\vecs_{4\ell+1}\bigr)$}, where two \hbox{$S\!=\!1/2$} spins
($\vecs_{4\ell-3}$ and $\vecs_{4\ell-2}$) and two \hbox{$S\!=\!1$} spins
($\vecS_{4\ell-1}$ and $\vecS_{4\ell}$) are arranged alternatively.  In
several limiting cases of $J_1$, $J_2$, and $J_3$ we apply the Wigner-Eckart
theorem and carry out a perturbation calculation to examine the behavior of
the massless lines where the energy gap vanishes.  Performing a quantum Monte
Carlo calculation without global flips at a sufficiently low temperature for
the case where $J_1\!=\!J_3\!=\!1.0$ and $J_2\!>\!0$, we find that the
ground state of the present system in this case undergoes a second-order phase
transition accompanying the vanishing of the energy gap at
\hbox{$J_2\!=\!J_{2{\rm c}}$ with $J_{2{\rm c}}\!=\!0.77\pm0.01$}.  We also
find that the ground states for both \hbox{$J_2\!<\!J_{2{\rm c}}$} and
\hbox{$J_2\!>\!J_{2{\rm c}}$} can be understood by means of the
valence-bond-solid picture.  A quantum Monte Carlo calculation which takes the
global flips along the Trotter direction into account is carried out to
elucidate the temperature dependences of the specific heat and the magnetic
susceptibility.  In particular, it is found that the susceptibility per unit
cell for $J_2\!=\!0.77$ with $J_1\!=\!J_3\!=\!1.0$ takes a finite value at
absolute zero temperature and that the specific heat per unit cell versus
temperature curve for $J_2\!=\!5.0$ with $J_1\!=\!J_3\!=\!1.0$ has a double
peak.
}

\kword
{
quantum mixed spin-1/2$\,$-1/2$\,$-1$\,$-1 chain, ground-state properties,
thermodynamic properties, quantum Monte Carlo calculation, ground-state
phase transition, specific heat, magnetic susceptibility
}

\begin{document}
\sloppy
\maketitle

\section{Introduction}

A quantum antiferromagnetic chain has been the subject of numerous
experimental as well as theoretical studies over a long period, in particular
since Haldane's prediction~\cite{rf:1} implying that the chain with an
integer spin is quite different from the chain with a half-integer spin in the
ground-state and low-lying-excited-state properties.  Almost all results of
these studies support the prediction, and now it is widely agreed that for the
case of isotropic nearest-neighbor exchange interactions, the former chain has
massive excitations with a exponential decay of the two-spin correlation
functions, while the latter chain has massless excitations with a power-law
decay of the two-spin correlation functions.~\cite{rf:2}

Under these circumstances, a mixed spin chain consisting of two kinds of
quantum spins which are arranged periodically has been of considerable interest
in recent years.  On the experimental side, several real materials of this
type have been synthesized, and their magnetic properties have been
observed.~\cite{rf:3}  A typical example of these materials is
CuNi(pba)(H$_2$O)$_3$$\cdot$2H$_2$O with
pba=1,3-propylenebis(oxamato),~\cite{rf:4} where an $S\!=\!1/2$ spin of the
Cu$^{2+}$ ion and an $S\!=\!1$ spin of the Ni$^{2+}$ ion make an alternating
array and each nearest-neighbor pair of the spins couples by the isotropic,
antiferromagnetic exchange interaction.  On the theoretical
side, on the other hand, a variety of models describing the quantum mixed spin
chain have been investigated.~\cite{rf:5,rf:6,rf:7,rf:8,rf:9,rf:10,rf:11,rf:12,rf:13,rf:14,rf:15,rf:16}  Integrable models have been constructed and
discussed by several authors.~\cite{rf:5,rf:6,rf:7,rf:8}  Although
these integrable models are specific ones in which rather complicated
interactions are assumed, exact solutions for the models are helpful for us
to understand the essential consequences of the quantum mixed spin
chain.  Motivated by the above experimental observations, several
authors~\cite{rf:9,rf:10,rf:11,rf:12,rf:14,rf:15,rf:16} have studied both
analytically and numerically a simpler case of the $S\!=\!1/2$ and $S\!=\!1$
alternating spin chain with antiferromagnetic nearest-neighbor exchange
interactions.  A characteristic feature of this chain in the case of isotropic
or Ising-type interactions is that it is a quantum ferrimagnet and its ground
state is magnetic.

Generally speaking, quantum fluctuations play a more crucial role in the
nonmagnetic ground state than in the magnetic one.  As an example of the
quantum mixed spin chains which have isotropic nearest-neighbor exchange
interactions only and also have a nonmagnetic ground state, we consider in
this paper a one-dimensional antiferromagnet where two $S\!=\!1/2$ and two
$S\!=\!1$ spins are arranged alternatively.  For the sake of simplicity we
call this system the spin-1/2$\,$-1/2$\,$-1$\,$-1 chain.  The Hamiltonian
is given by

\begin{full}
\begin{equation}
\label{eq:1.1}
    {\cal H} = \sum_{\ell=1}^{N/4}
          \Bigl(J_1\,\vecs_{4\ell-3}\cdot\vecs_{4\ell-2}
              + J_2\,\vecs_{4\ell-2}\cdot\vecS_{4\ell-1}
              + J_3\,\vecS_{4\ell-1}\cdot\vecS_{4\ell}
              + J_2\,\vecS_{4\ell  }\cdot\vecs_{4\ell+1}\Bigr)\,,
\end{equation}
\end{full}

\noindent
where $\vecs_{\ell}$ with $\ell\!=\!1$ and $2$ (mod $4$) and $\vecS_{\ell}$
with $\ell\!=\!3$ and $4$ (mod $4$) are, respectively, the $S\!=\!1/2$ and
$S\!=\!1$ spin operators at the $\ell$th site; $J_1$, $J_2$, and $J_3$ are,
respectively, the interaction constants between the nearest-neighbor pair of
$S\!=\!1/2$ spins, that of $S\!=\!1/2$ and $S\!=\!1$ spins, and that of
$S\!=\!1$ spins; $N$, being assumed to be a multiple of four, is the total
number of spins.  We impose periodic boundary conditions
($\vecs_{N+1}\!\equiv\!\vecs_{1}$).

It is naturally anticipated that the present mixed spin system has different
ground states depending upon the signs of $J_1$, $J_2$, and $J_3$.  As we will
discuss later in more detail (see Table~\ref{table:1} below), the value
${\cal S}_{\rm tot}$ of the total spin
\begin{equation}
 \veccalS_{\rm tot} = 
     \sum_{\ell=1}^{N/4}\bigl\{
           \vecs_{4\ell-3} + \vecs_{4\ell-2} + \vecS_{4\ell-1}
         + \vecS_{4\ell}\bigr\}
\end{equation}
in the ground state can be determined rigorously by applying the Lieb-Mattis
theorem.~\cite{rf:17}  Except in the region where \hbox{$J_1\!<\!0$} and
\hbox{$J_3\!<\!0$}, the ground state belongs to the
\hbox{${\cal S}_{\rm tot}\!=\!0$}
subspace, and therefore it is nonmagnetic.  Furthermore, in the region where
\hbox{$J_1\!<\!0$}, \hbox{$J_2\!>\!0$}, and \hbox{$J_3\!<\!0$} and in the
region where \hbox{$J_1\!<\!0$}, \hbox{$J_2\!<\!0$}, and \hbox{$J_3\!<\!0$},
the ground states belong, respectively, to the
${\cal S}_{\rm tot}\!=\!N/4$ and ${\cal S}_{\rm tot}\!=\!3N/4$ subspaces.  The
former is ferrimagnetic, while the latter is ferromagnetic.  For comparison
we depict in Fig.~\ref{fig:1} the spin array with the lowest energy of
the corresponding classical system in which $\vecs_\ell$ and $\vecS_\ell$ are
replaced, respectively, by the classical spin vectors whose magnitudes are
$1/2$ and $1$.  From this figure and Table~\ref{table:1} we find that there is
a close analogy between the classical lowest-energy state and the quantum
ground state.  It is interesting to see that the periods concerning the
translational symmetry of the classical lowest-energy states are,
respectively, four and eight in units of the lattice spacing in the region
where \hbox{$J_1J_3\!>\!0$} and in the region where \hbox{$J_1J_3\!<\!0$}.

We explore here both analytically and numerically the ground-state and
thermodynamic properties of the present system.~\cite{rf:18}  In several
limiting cases of the interaction constants we apply the Wigner-Eckart
theorem~\cite{rf:19} and carry out a perturbation calculation to examine the
behavior of the massless lines in the $J_1$ versus $J_2$ plane with
\hbox{$J_1\!>\!0$} and \hbox{$J_3\!>\!0$}.  Confining ourselves to the case
where $J_1\!=\!J_3\!=\!1.0$ and $J_2\!>\!0$, we perform numerical
calculations.  In these calculations we mainly employ a quantum Monte
Carlo (QMC) method, and do an exact-diagonalization calculation for $N\!=\!8$
and $16$ to check the QMC results.  To determine as accurately as possible the
massless point $J_{2{\rm c}}$ of $J_2$ in the above case is one of our main
purposes.  We discuss how the ground states for both $J_2\!<\!J_{2{\rm c}}$
and $J_2\!>\!J_{2{\rm c}}$ are represented in the frame of the
valence-bond-solid (VBS) picture proposed by Affleck, Kennedy, Lieb, and
Tasaki.~\cite{rf:20}  We also aim at clarifying the temperature dependences of
the specific heat and the magnetic susceptibility for a few values of $J_2$ in
this case.  Very recently, Fukui and Kawakami~\cite{rf:14} have also discussed
this system.  They have mapped this system to the non-linear sigma model and
have shown that in its ground state a second-order phase transition which
accompanies the vanishing of the energy gap may occur.

The remainder of the paper is organized as follows.  The next section
(\S2) and the following one (\S3) are devoted to analytical treatments and
numerical calculations, respectively.  Finally, the main results are
summarized and further discussed in \S4.

\section{Analytical Treatments}

We start with applying the Lieb-Mattis theorem~\cite{rf:17} to determine the
value ${\cal S}_{\rm tot}$ of the total spin $\veccalS_{\rm tot}$ in the
ground state for arbitrary values of $J_1$, $J_2$, and $J_3$.  In the present
case the theorem implies that, if we can divide the lattice into the $A$ and
$B$ sublattices in such a way that all the intrasublattice interaction
constants are not positive and all the intersublattice interaction constants
are not negative, then ${\cal S}_{\rm tot}$ is given by 
\begin{equation}
   {\cal S}_{\rm tot} 
      = \big\vert {\cal S}_{{\rm max},A} - {\cal S}_{{\rm max},B} \big\vert\,,
\end{equation}
where ${\cal S}_{{\rm max},A}$ and ${\cal S}_{{\rm max},B}$ are, respectively,
the maximum values of the total spins in the $A$ and $B$ sublattices.  (Note
that the constant $g^2$ in ref.$\,$17 is equal to zero because the
Hamiltonian ${\cal H}$ consists of the nearest-neighbor interactions
only.)  Let us consider two examples.  For the region where \hbox{$J_1\!>\!0$},
\hbox{$J_2\!>\!0$}, and \hbox{$J_3\!>\!0$}, we choose the sites $\ell_A$
belonging to the $A$ sublattice to be $\ell_A\!=\!1$ (mod $2$) and the sites
$\ell_B$ belonging to the $B$ sublattice to be $\ell_B\!=\!2$ (mod $2$).  Then,
all of $J_1$, $J_2$, and $J_3$ are the intersublattice interaction constants,
which satisfy the above requirement, and thus we obtain
\hbox{${\cal S}_{\rm tot}\!=\!0$} since 
\hbox{${\cal S}_{{\rm max},A}\!=\!{\cal S}_{{\rm max},B}\!=\!3N/8$}.  For the
region where \hbox{$J_1\!>\!0$}, \hbox{$J_2\!<\!0$}, and \hbox{$J_3\!>\!0$},
on the other hand, we choose $\ell_A$ to be $\ell_A\!=\!1$ and
$4$ (mod $4$) and $\ell_B$ to be $\ell_B\!=\!2$ and $3$ (mod $4$).  Then,
$J_2$ is the intrasublattice interaction constant and the remaining $J_1$ and
$J_3$ are the intersublattice interaction constants, which again satisfy
the above requirement.  Thus, we obtain again ${\cal S}_{\rm tot}\!=\!0$ since
${\cal S}_{{\rm max},A}\!=\!{\cal S}_{{\rm max},B}\!=\!3N/8$.  In a similar
way, we can determine ${\cal S}_{\rm tot}$ for the remaining regions of $J_1$,
$J_2$, and $J_3$.  The results are listed in Table~\ref{table:1} together
with $\ell_A$, $\ell_B$, ${\cal S}_{{\rm max},A}$, and
${\cal S}_{{\rm max},B}$ for each region.  It should be noted that the $A$
and $B$ sublattices correspond, respectively, to the ^^ up' and ^^ down'
sublattices in the lowest-energy classical-spin arrays shown in
Fig.~\ref{fig:1}.

\begin{fulltable}
\caption{Values of $\ell_A$, $\ell_B$, ${\cal S}_{{\rm max},A}$,
${\cal S}_{{\rm max},B}$, and ${\cal S}_{\rm tot}$ for each region of $J_1$,
$J_2$, and $J_3$.}
\label{table:1}
\begin{fulltabular}{@{\hspace{\tabcolsep}\extracolsep{\fill}}cccccc}
  \hline
  Region of $J_1$, $J_2$, and $J_3$ & $\ell_A$ & $\ell_B$ &
  ${\cal S}_{{\rm max},A}$ & ${\cal S}_{{\rm max},B}$ & ${\cal S}_{\rm tot}$
  \\
  \hline
  $J_1\!>0$, $J_2\!>\!0$, $J_3\!>\!0$ & $1$ (mod $2$) & $2$ (mod $2$) &
  $3N/8$ & $3N/8$ & $0$ \\
  $J_1\!>0$, $J_2\!<\!0$, $J_3\!>\!0$ & $1$, $4$ (mod $4$) &
  $2$, $4$ (mod $4$) & $3N/8$ & $3N/8$ & $0$ \\
  $J_1\!<0$, $J_2\!<\!0$, $J_3\!>\!0$ & $1$, $2$, $3$, $8$ (mod $8$) &
  $4$, $5$, $6$, $7$ (mod $8$) & $3N/8$ & $3N/8$ & $0$ \\
  $J_1\!<0$, $J_2\!>\!0$, $J_3\!>\!0$ & $1$, $2$, $4$, $7$ (mod $8$) &
  $3$, $5$, $6$, $8$ (mod $8$) & $3N/8$ & $3N/8$ & $0$ \\
  $J_1\!>0$, $J_2\!>\!0$, $J_3\!<\!0$ & $1$, $3$, $4$, $6$ (mod $8$) &
  $2$, $5$, $7$, $8$ (mod $8$) & $3N/8$ & $3N/8$ & $0$ \\
  $J_1\!>0$, $J_2\!<\!0$, $J_3\!<\!0$ & $1$, $6$, $7$, $8$ (mod $8$) &
  $2$, $3$, $4$, $5$ (mod $8$) & $3N/8$ & $3N/8$ & $0$ \\
  $J_1\!<0$, $J_2\!<\!0$, $J_3\!<\!0$ & all & none & $3N/4$ & 0 & $3N/4$ \\
  $J_1\!<0$, $J_2\!>\!0$, $J_3\!<\!0$ & $1$, $2$ (mod $4$) &
  $3$, $4$ (mod $4$) & $N/4$ & $N/2$ & $N/4$ \\
\hline
\end{fulltabular}
\end{fulltable}

We now turn to an application of the Wigner-Eckart theorem~\cite{rf:19} to
the following three limiting cases:~(a) the case where
\hbox{$\vert J_2\vert\!\gg\!\vert J_1\vert,\;\vert J_3\vert$}
with \hbox{$J_2\!>\!0$}, (b) the case where
\hbox{$\vert J_2\vert\!\gg\!\vert J_1\vert,\;\vert J_3\vert$}
with \hbox{$J_2\!<\!0$}, and (c) the case where
\hbox{$\vert J_1\vert\!\gg\!\vert J_2\vert,\;\vert J_3\vert$}
with \hbox{$J_1\!<\!0$}.  In
the limiting case (a), a pair of the spins $\vecs_{4\ell-2}$ and
$\vecS_{4\ell-1}$ can be replaced by an effective $S\!=\!1/2$ spin
$\vecs^{\;\rm eff}_{4\ell-2,4\ell-1}$.  This replacement means that we take
into account only the doublet state of the two-spin system consisting of
$\vecs_{4\ell-2}$ and $\vecS_{4\ell-1}$ and neglect the quartet state of the
system.  Similarly, a pair of the spins $\vecS_{4\ell}$ and $\vecs_{4\ell+1}$
can be replaced by an effective $S\!=\!1/2$ spin
$\vecs^{\;\rm eff}_{4\ell,4\ell+1}$  Then, the Wigner-Eckart theorem implies
that both $\vecs_{4\ell-2}$ and $\vecS_{4\ell-1}$ are proportional to
$\vecs^{\;\rm eff}_{4\ell-2,4\ell-1}$ and both $\vecS_{4\ell}$ and
$\vecs_{4\ell+1}$ are proportional to $\vecs^{\;\rm eff}_{4\ell,4\ell+1}$:
\begin{subeqnarray}
    \vecs_{4\ell-2} = \alpha\,\vecs^{\;\rm eff}_{4\ell-2,4\ell-1}\,,  \\
    \vecS_{4\ell-1} = \beta\,\vecs^{\;\rm eff}_{4\ell-2,4\ell-1}\,,   \\
    \vecS_{4\ell}   = \beta'\,\vecs^{\;\rm eff}_{4\ell,4\ell+1}\,,    \\
    \vecs_{4\ell+1} = \alpha'\,\vecs^{\;\rm eff}_{4\ell,4\ell+1}\,,
\end{subeqnarray}
$\alpha$, $\beta$, $\beta'$, and $\alpha'$ being constants.  The values of
$\alpha$, $\beta$, $\beta'$, and $\alpha'$ can be obtained, respectively, by
calculating all matrix elements of both sides of eq.$\,$(2.2a), eq.$\,$(2.2b),
eq.$\,$(2.2c), and eq.$\,$(2.2d) in the doublet subspace.  The results are
$\alpha\!=\!\alpha'\!=\!-1/3$ and $\beta\!=\!\beta'\!=\!4/3$.  Substituting
eqs.$\,$(2.2a)-(2.2d) with these constants into eq.$\,$(1.1), we obtain,
apart from the constant term coming from the $J_2$-terms,

\begin{full}
\begin{equation}
\label{eq:2.3}
    {\cal H_{({\rm a})}} = \sum_{\ell=1}^{N/4}
          \biggl(\frac{1}{9} J_1\,\vecs^{\;\rm eff}_{4\ell-4,4\ell-3}\cdot
                                 \vecs^{\;\rm eff}_{4\ell-2,4\ell-1}
              + \frac{16}{9}J_3\,\vecs^{\;\rm eff}_{4\ell-2,4\ell-1}\cdot
                                 \vecs^{\;\rm eff}_{4\ell,4\ell+1}\biggr)
\end{equation}
\end{full}

\noindent
with $\vecs^{\;\rm eff}_{0,1}\!\equiv\!\vecs^{\;\rm eff}_{N,N+1}\!\equiv\!\vecs^{\;\rm eff}_{N,1}$.  Thus, we can conclude that in the limiting
case (a), our system described by the Hamiltonian ${\cal H}$ of
eq.$\,$(1.1) is equivalent to the $S\!=\!1/2$ bond-alternating chain
described by the Hamiltonian ${\cal H_{({\rm a})}}$ of eq.$\,$(2.3), as
far as the ground state and the sufficiently low-energy excited states are
concerned.

In the limiting case (b), we replace a pair of the spins $\vecs_{4\ell-2}$ and
$\vecS_{4\ell-1}$ ($\vecS_{4\ell}$ and $\vecs_{4\ell+1}$) by an effective
$S\!=\!3/2$ spin $\vecSigma^{\;\rm eff}_{4\ell-2,4\ell-1}$
($\vecSigma^{\;\rm eff}_{4\ell,4\ell+1}$), neglecting the doublet state of the
two-spin system consisting of $\vecs_{4\ell-2}$ and $\vecS_{4\ell-1}$
($\vecS_{4\ell}$ and $\vecs_{4\ell+1}$).  Applying the Wigner-Eckart theorem,
we obtain
\begin{subeqnarray}
  \vecs_{4\ell-2} = \frac{1}{3}\vecSigma^{\;\rm eff}_{4\ell-2,4\ell-1}\,,  \\
  \vecS_{4\ell-1} = \frac{2}{3}\vecSigma^{\;\rm eff}_{4\ell-2,4\ell-1}\,,  \\
  \vecS_{4\ell}   = \frac{2}{3}\vecSigma^{\;\rm eff}_{4\ell,4\ell+1}\,,    \\
  \vecs_{4\ell+1} = \frac{1}{3}\vecSigma^{\;\rm eff}_{4\ell,4\ell+1}\,.
\end{subeqnarray}
Thus, we can show that in the limiting case (b), our system is equivalent to
the $S\!=\!3/2$ bond-alternating chain described by the following Hamiltonian:

\begin{full}
\begin{equation}
\label{eq:2.4}
  {\cal H_{({\rm b})}} = \sum_{\ell=1}^{N/4}
       \biggl(\frac{1}{9} J_1\,\vecSigma^{\;\rm eff}_{4\ell-4,4\ell-3}\cdot
                               \vecSigma^{\;\rm eff}_{4\ell-2,4\ell-1}
             + \frac{4}{9}J_3\,\vecSigma^{\;\rm eff}_{4\ell-2,4\ell-1}\cdot
                               \vecSigma^{\;\rm eff}_{4\ell,4\ell+1}\biggr)
\end{equation}
\end{full}

\noindent
with $\vecSigma^{\;\rm eff}_{0,1}\!\equiv\!\vecSigma^{\;\rm eff}_{N,N+1}\!\equiv\!\vecSigma^{\;\rm eff}_{N,1}$, as far as the ground state and the
sufficiently low-energy excited states are concerned.  Furthermore, in the
limiting case (c), we replace a pair of the spins $\vecs_{4\ell-3}$ and
$\vecs_{4\ell-2}$ by an effective $S\!=\!1$ spin
$\vecS^{\;\rm eff}_{4\ell-3,4\ell-2}$, neglecting the singlet state of the
two-spin system consisting of $\vecs_{4\ell-3}$ and $\vecs_{4\ell-2}$.  The
Wigner-Eckart theorem gives
\begin{subeqnarray}
  \vecs_{4\ell-3} = \frac{1}{2}\vecS^{\;\rm eff}_{4\ell-3,4\ell-2}\,,  \\
  \vecs_{4\ell-2} = \frac{1}{2}\vecS^{\;\rm eff}_{4\ell-3,4\ell-2}\,.
\end{subeqnarray}
Thus, our system in the limiting case (c) is equivalent to the $S\!=\!1$
chain described by the following Hamiltonian:

\begin{full}
\begin{equation}
\label{eq:2.6}
  {\cal H_{({\rm c})}} = \sum_{\ell=1}^{N/4}
       \biggl(\frac{1}{2} J_2\,\vecS^{\;\rm eff}_{4\ell-3,4\ell-2}\cdot
                               \vecS_{4\ell-1}
                        + J_3\,\vecS_{4\ell-1}\cdot\vecS_{4\ell}
            + \frac{1}{2} J_2\,\vecS_{4\ell}\cdot
                               \vecS^{\;\rm eff}_{4\ell+1,4\ell+2}\biggr)
\end{equation}
\end{full}

\noindent
with $\vecS^{\;\rm eff}_{N+1,N+2}\!\equiv\!\vecS^{\;\rm eff}_{1,2}$, as far
as the ground state and the sufficiently low-energy excited states are
concerned.

From the above arguments for the first two limiting cases (a) and (b), we can
discuss when the system has massless excitations in these cases.  In the case
(a), it is massless only when
\begin{equation}
   J_1 = 16J_3 > 0\,,
\end{equation}
since the antiferromagnetic $S\!=\!1/2$ bond-alternating chain becomes
massless only in the case of no bond-alternation.~\cite{rf:21}  In the case
(b), on the other hand, the system is massless when~\cite{rf:22,rf:23}
\begin{equation}
   J_1 = 4J_3 > 0 \quad {\rm and} \quad
   J_1 = 4J_3 \frac{1 \pm \delta}{1 \mp \delta} > 0\,.
\end{equation}
The value of $\delta$ has been determined~\cite{rf:24,rf:25} recently to be
\hbox{$\delta\!=\!0.43\pm0.01$}, which is the value given in ref.$\,$25.

We also investigate the possibility of massless lines in the limiting
cases:~(d) the case where
\hbox{$J_3\!\gg\!\vert J_1\vert,\;\vert J_2\vert$} and (e) the case where
\hbox{$J_1\!\gg\!\vert J_2\vert$} and \hbox{$J_1\!\gg\!J_3\!>\!0$}.  For
this purpose we perform perturbation calculations.  For the case (d), we first
observe that the system is massless at the point \hbox{$J_1\!=\!J_2\!=\!0.0$}
when $J_3\!>\!0$.  This is because at this point the system consists of
$N/4$ independent pairs of two $S\!=\!1$ spins, each of which has a singlet
state as the lowest-energy state, plus $N/2$ free $S\!=\!1/2$ spins, and
therefore the ground state of the system is $2^{N/2}$-fold degenerate.  In
order to explore the neighborhood of the above point for finite $J_3(>\!0)$,
we consider the four-spin system described by the Hamiltonian,

\begin{full}
\begin{equation}
         h^{(4)}_\ell =   J_2\,\vecs_{4\ell-2}\cdot\vecS_{4\ell-1}
                        + J_3\,\vecS_{4\ell-1}\cdot\vecS_{4\ell}
                        + J_2\,\vecS_{4\ell  }\cdot\vecs_{4\ell+1}\,.
\end{equation}
\end{full}

\noindent
Diagonalizing analytically this Hamiltonian, we find that the ground and
first-excited states of the four-spin system are singlet and triplet states,
respectively.  The energy difference ${\tilde J}$ between them is given by

\begin{full}
\begin{equation}
 {\tilde J} = J_2 \biggl\{\frac{4}{3}\Bigl(\frac{J_2}{J_3}\Bigr)
                        + 2          \Bigl(\frac{J_2}{J_3}\Bigr)^2 
                        - \frac{47}{27}\Bigl(\frac{J_2}{J_3}\Bigr)^3
            + O\biggl(\Bigl(\frac{J_2}{J_3}\Bigr)^4\biggl)\biggr\}\,,
\end{equation}
\end{full}

\noindent
while the energies of the second- and higher-excited states measured from the
ground-state energy are finite at $J_2\!=\!0.0$.  Thus, taking only the ground
and first-excited states into consideration, we regard the four-spin system as
the system described by ${\tilde J}\vecsigma_\ell\cdot\vectau_\ell$, where
both $\vecsigma_\ell$ and $\vectau_\ell$ are $S\!=\!1/2$ operators.  Then,
as far as the ground state and the sufficiently low-energy
excited states are concerned, we map our original system described by the
Hamiltonian ${\cal H}$ of eq.$\,$(1.1) to the following effective Hamiltonian:

\begin{full}
\begin{equation}
  {\cal H}_{\rm eff} = \sum_{\ell=1}^{N/4}\Bigl(
         {\tilde J}   \vecsigma_\ell\cdot\vectau_\ell
       + {\tilde J}_1 \vectau_\ell  \cdot\vecsigma_{\ell+1}
       + {\tilde J}_2 \vecsigma_\ell\cdot\vecsigma_{\ell+1}
       + {\tilde J}_2 \vectau_\ell  \cdot\vectau_{\ell+1}\Bigr)
\end{equation}
\end{full}

\noindent
with $\vecsigma_{\frac{N}{4}+1}\!\equiv\!\vecsigma_{1}$ and
$\vectau_{\frac{N}{4}+1}\!\equiv\!\vectau_{1}$.  The values of ${\tilde J}_1$
and ${\tilde J}_2$ are determined in such a way that the matrix elements of
the two operators ${\tilde J}_1\vectau_\ell\cdot\vecsigma_{\ell+1}\!+\!{\tilde J}_2\vecsigma_\ell\cdot\vecsigma_{\ell+1}\!+\!{\tilde J}_2 \vectau_\ell\cdot\vectau_{\ell+1}$ and $J_1\vecs_{4\ell+1}\cdot\vecs_{4\ell+2}$ coincide.  The
mapping is correct to the order of $(J_2/J_3)^3$, and ${\tilde J}_1$ and
${\tilde J}_2$ are given by

\begin{full}
\begin{subeqnarray}
 {\tilde J}_1 &&\hspace{-0.80truecm}
   = J_1 \biggl\{1 - \frac{4}{3}\Bigl(\frac{J_2}{J_3}\Bigr)^2
                   - \frac{4}{3}\Bigl(\frac{J_2}{J_3}\Bigr)^3
                   + O\biggl(\Bigl(\frac{J_2}{J_3}\Bigr)^4\biggl)\biggr\}\,, \\
 {\tilde J}_2 &&\hspace{-0.80truecm}
   = J_1 \biggl\{\frac {2}{3} \Bigl(\frac{J_2}{J_3}\Bigr)^3
                   + O\biggl(\Bigl(\frac{J_2}{J_3}\Bigr)^4\biggl)\biggr\}\,.
\end{subeqnarray}
\end{full}

\noindent
Thus, to the order of $(J_2/J_3)^3$, the effective Hamiltonian
${\cal H}_{\rm eff}$ describes the $S\!=\!1/2$ chain which has the
bond-alternating nearest-neighbor interactions with the interaction constants
${\tilde J}$ and ${\tilde J}_1$ and also the uniform next-nearest-neighbor
interaction with the interaction constant ${\tilde J}_2$.  Thus, remembering
the fact that the antiferromagnetic $S\!=\!1/2$ chain with the uniform
nearest-neighbor and next-nearest-neighbor interactions where the ratio of the
latter interaction to the former one is less than 0.24 is
massless,~\cite{rf:26} we may conclude, from eqs.$\,$(2.11), (2.13a), and
(2.13b), that in the limiting case (d), the present quantum mixed spin chain
described by the Hamiltonian ${\cal H}$ of eq.$\,$(1.1) is massless when

\begin{full}
\begin{equation}
   J_1 = J_2 \biggl\{\frac{4}{3}\Bigl(\frac{J_2}{J_3}\Bigr)
                   + 2          \Bigl(\frac{J_2}{J_3}\Bigr)^2
                   + \frac{1}{27}\Bigl(\frac{J_2}{J_3}\Bigr)^3
       + O\biggl(\Bigl(\frac{J_2}{J_3}\Bigr)^4\biggl)\biggl\}\,.
\end{equation}
\end{full}

For the case (e) we perform a similar perturbation calculation in the following
way.  We first diagonalize analytically the Hamiltonian which describes
another four-spin system,

\begin{full}
\begin{equation}
         h^{(4)}_\ell =   J_2\,\vecS_{4\ell-4}\cdot\vecs_{4\ell-3}
                        + J_1\,\vecs_{4\ell-3}\cdot\vecs_{4\ell-2}
                        + J_2\,\vecs_{4\ell-2}\cdot\vecS_{4\ell-1}\,,
\end{equation}
\end{full}

\noindent
to obtain the energy difference ${\tilde J}$ between the singlet ground and
triplet first-excited states and that ${\tilde J}'$ between the singlet ground
and quintet second-excited state to be
\begin{subeqnarray}
 {\tilde J}  = &&\hspace{-0.80truecm}
            J_1 \biggl\{\frac{1}{2}\Bigl(\frac{J_2}{J_1}\Bigr)^2 
          + O\biggl(\Bigl(\frac{J_2}{J_1}\Bigr)^3\biggl)\biggr\}\,,  \\
 {\tilde J}' = &&\hspace{-0.80truecm}
            J_1 \biggl\{\frac{3}{2}\Bigl(\frac{J_2}{J_1}\Bigr)^2 
          + O\biggl(\Bigl(\frac{J_2}{J_1}\Bigr)^3\biggl)\biggr\}\,.
\end{subeqnarray}
Adding the fact that the energy difference between these lowest states and
higher-excited states are finite at \hbox{$\vert J_2\vert/J_1\!=\!0.0$}, we
can regard, to the order of $(J_2/J_1)^2$, the four-spin system as a system
described by ${\tilde J}\vecsigma_\ell\cdot\vectau_\ell$, where
$\vecsigma_\ell$ and $\vectau_\ell$ are $S\!=\!1$ operators.  Then, as far as
the ground state and the sufficiently low-energy excited states are concerned,
we map the present system described by the Hamiltonian ${\cal H}$ of
eq.$\,$(1.1) to the effective Hamiltonian of eq.$\,$(2.12) with the $S\!=\!1$
operators $\vecsigma_\ell$ and $\vectau_\ell$.  The values of ${\tilde J}_1$
and ${\tilde J}_2$ can be determined similarly to the above
perturbation calculation.  The results are given by
\begin{subeqnarray}
  {\tilde J}_1 &&\hspace{-0.80truecm}
     = J_3 \biggl\{1 - \frac{1}{2}\Bigl(\frac{J_2}{J_1}\Bigr)^2 
     + O\biggl(\Bigl(\frac{J_2}{J_1}\Bigr)^3\biggl)\biggl\}\,,     \\
  {\tilde J}_2 &&\hspace{-0.80truecm}
     = J_3 \;O\biggl(\Bigl(\frac{J_2}{J_1}\Bigr)^3\biggl)\,,
\end{subeqnarray}
which show that, to the order of $(J_2/J_1)^2$, the effective
Hamiltonian ${\cal H}_{\rm eff}$ describes the $S\!=\!1$ chain which has the
bond-alternating nearest-neighbor interactions with the interaction constants
${\tilde J}$ and ${\tilde J}_1$.  Thus, it may be concluded, from
eqs.$\,$(2.16a) and (2.17a), that in the limiting case (e), the present
system is massless when
$(J_1/2)(J_2/J_1)^2\big/J_3\bigl\{1-(1/2)(J_2/J_1)^2\bigr\}\!=\!\bigl(1\pm\delta\bigr)\big/\bigl(1\mp\delta\bigr)$ with
$\delta\!\simeq\!1/4$,~\cite{rf:25,rf:27,rf:28,rf:29,rf:30} or
equivalently, when
\begin{equation}
   J_1 = \frac{3}{10} \frac{J_2^2}{J_3} \quad {\rm and} \quad 
   J_1 = \frac{5}{6}  \frac{J_2^2}{J_3}\,.
\end{equation}

The above results given by eqs.$\,$(2.8), (2.9), (2.14), and (2.18) with $J_3$
fixed at a positive and finite value describe, respectively, the asymptotic
behavior in the limiting cases (a), (b), (d), and (e) of the massless lines on
the $J_1$ versus $J_2$ plane.  These are summarized in Fig.~\ref{fig:2}, where
$J_3$ is chosen to be $J_3\!=\!1.0$.  Our preliminary results indicate that the
asymptotic massless lines in Fig.~\ref{fig:2} connect to form four massless
lines.  Leaving the detailed presentation of these massless lines for further
publication,~\cite{rf:31} we determine in the next section the massless point
of $J_2$ for the case where $J_1\!=\!J_3\!=\!1.0$ and $J_2\!>\!0$, as
mentioned in \S1.

\section{Numerical calculations}

\subsection{QMC method}

As has been mentioned in \S1, we mainly employ a QMC method in the numerical
calculation.  Before discussing the numerical results, we explain the QMC
procedure for the present spin-1/2$\,$-1/2$\,$-1$\,$-1 chain.  By the use of
the Suzuki-Trotter decomposition~\cite{rf:32} of checker-board
type,~\cite{rf:33} the partition function
$Z\!=\!{\rm Tr}\bigl[\exp(-\beta{\cal H})\bigr]$, where
$\beta\!=\!(k_{\rm B}T)^{-1}$ with the Boltzmann constant $k_{\rm B}$ and the
absolute temperature $T$, is approximated as

\begin{full}
\begin{eqnarray}
\label{eq:3.1}
   Z \simeq {\rm Tr}\Biggl[\prod_{\ell=1}^{N/4}&&
              \Bigl\{\exp\bigl(-\beta h_{4\ell-3,4\ell-2}/n\bigr)
                     \exp\bigl(-\beta h_{4\ell-1,4\ell}  /n\bigr)\Bigr\}^n
                                                              \nonumber    \\
                &&\times  \prod_{\ell=1}^{N/4}
              \Bigl\{\exp\bigl(-\beta h_{4\ell-2,4\ell-1}/n\bigr)
                     \exp\bigl(-\beta h_{4\ell,  4\ell+1}/n\bigr)\Bigr\}^n
           \Biggr]\,.
\end{eqnarray}
\end{full}

\noindent
Here,
\begin{subeqnarray}
        h_{4\ell-3,4\ell-2} = J_1\,\vecs_{4\ell-3}\cdot\vecs_{4\ell-2}\,,  \\
        h_{4\ell-1,4\ell}   = J_3\,\vecS_{4\ell-1}\cdot\vecS_{4\ell}\,,    \\
        h_{4\ell-2,4\ell-1} = J_2\,\vecs_{4\ell-2}\cdot\vecS_{4\ell-1}\,,  \\
        h_{4\ell,  4\ell+1} = J_2\,\vecS_{4\ell  }\cdot\vecs_{4\ell+1}
\end{subeqnarray}
are the local Hamiltonians, and $n$ is the Trotter number.  Choosing the
$z$-axis as the quantization axis, we introduce the local Boltzmann factors
defined by

\begin{full}
\begin{subeqnarray}
 \rho_{4\ell-3,4\ell-2}^{(2r-1,2r)}
    = \Bigl\langle s_{4\ell-3}^{(2r-1)}s_{4\ell-2}^{(2r-1)}\Bigl\vert
           \exp\bigl(-\beta h_{4\ell-3,4\ell-2}/n\bigr)
      \Bigr\vert   s_{4\ell-3}^{(2r  )}s_{4\ell-2}^{(2r  )}\Bigr\rangle\,, \\
 \rho_{4\ell-1,4\ell  }^{(2r-1,2r)}
    = \Bigl\langle S_{4\ell-1}^{(2r-1)}S_{4\ell  }^{(2r-1)}\Bigl\vert
           \exp\bigl(-\beta h_{4\ell-1,4\ell  }/n\bigr)
      \Bigr\vert   S_{4\ell-1}^{(2r  )}S_{4\ell  }^{(2r  )}\Bigr\rangle\,, \\
 \rho_{4\ell-2,4\ell-1}^{(2r  ,2r+1)}
    = \Bigl\langle s_{4\ell-2}^{(2r  )}S_{4\ell-1}^{(2r  )}\Bigl\vert
           \exp\bigl(-\beta h_{4\ell-2,4\ell-1}/n\bigr)
      \Bigr\vert   s_{4\ell-2}^{(2r+1)}S_{4\ell-1}^{(2r+1)}\Bigr\rangle\,, \\
 \rho_{4\ell  ,4\ell+1}^{(2r  ,2r+1)}
    = \Bigl\langle S_{4\ell  }^{(2r  )}s_{4\ell+1}^{(2r  )}\Bigl\vert
           \exp\bigl(-\beta h_{4\ell  ,4\ell+1}/n\bigr)
      \Bigr\vert   S_{4\ell  }^{(2r+1)}s_{4\ell+1}^{(2r+1)}\Bigr\rangle\,,
\end{subeqnarray}
\end{full}

\noindent
with $\big\vert s_{4\ell-3}^{(r')}s_{4\ell-2}^{(r')}\big\rangle$,
$\big\vert S_{4\ell-1}^{(r')}S_{4\ell}^{(r')}\big\rangle$,
$\big\vert s_{4\ell-2}^{(r')}S_{4\ell-1}^{(r')}\big\rangle$, and
$\big\vert S_{4\ell}^{(r')}s_{4\ell+1}^{(r')}\big\rangle$, which denote,
respectively, the spin states consisting
of $\vecs_{4\ell-3}$ and $\vecs_{4\ell-2}$,
of $\vecS_{4\ell-1}$ and $\vecS_{4\ell}$,
of $\vecs_{4\ell-2}$ and $\vecS_{4\ell-1}$,
and of $\vecS_{4\ell}$ and $\vecs_{4\ell+1}$.  Here, $r'$ is a label along the
Trotter direction, taking the values, $1$, $2$, $\cdots$, $2n$;
$s_{4\ell-3}^{(r')}$ and $s_{4\ell-2}^{(r')}$ take the values,
$\pm\frac{1}{2}$; $S_{4\ell-1}^{(r')}$ and $S_{4\ell}^{(r')}$ take the values,
$0$, $\pm 1$.  Using these local Boltzmann factors, eq.$\,$(3.1) is rewritten
as

\begin{full}
\begin{equation}
\label{eq:3.4}
  Z \simeq \sum_{\{s_{4\ell-3}^{(r')}\}} \sum_{\{s_{4\ell-2}^{(r')}\}}
           \sum_{\{S_{4\ell-1}^{(r')}\}} \sum_{\{S_{4\ell  }^{(r')}\}}
           \prod_{r=1}^n \Biggl\{
           \prod_{\ell=1}^{N/4} \Bigl(\rho_{4\ell-3,4\ell-2}^{(2r-1,2r)}\;
                                \rho_{4\ell-1,4\ell  }^{(2r-1,2r)}\Bigr)
           \prod_{\ell=1}^{N/4} \Bigl(\rho_{4\ell-2,4\ell-1}^{(2r  ,2r+1)}\;
                                \rho_{4\ell  ,4\ell+1}^{(2r  ,2r+1)}\Bigr)
           \Biggr\}\,,
\end{equation}
\end{full}

\noindent
where
$\rho_{4\ell-2,4\ell-1}^{(2n,2n+1)}\!\equiv\!\rho_{4\ell-2,4\ell-1}^{(2n,1)}$
and
$\rho_{4\ell,4\ell+1}^{(2n,2n+1)}\!\equiv\rho_{4\ell,4\ell+1}^{(2n,1)}$.  The
right-hand side of eq.$\,$(3.4) can be regarded as the partition
function of a two-dimensional Ising system with the Ising
variables, $\{s_{4\ell-3}^{(r')}\}$, $\{s_{4\ell-2}^{(r')}\}$,
$\{S_{4\ell-1}^{(r')}\}$, and $\{S_{4\ell}^{(r')}\}$, which has four-body
interactions corresponding to the local Boltzmann factors.  A graphical
representation of this two-dimensional Ising system is presented in
Fig.~\ref{fig:3}.  It is noted that in this figure we have four kinds of
plaquettes, corresponding to four kinds of the local Boltzmann factors.

We perform a QMC calculation on the basis of the two-dimensional Ising
system discussed above.  To update the spin configuration, we carry out the
following local flips, keeping the total magnetization of the system
constant:~\cite{rf:34}

\begin{full}
\begin{subeqnarray}
\label{eq:3.5}
    \Bigl\{
       s_{4\ell-3}^{(2r)  }= \frac{1}{2},\,&&\hspace{-0.80truecm}
       s_{4\ell-2}^{(2r)  }=-\frac{1}{2},\,
       s_{4\ell-3}^{(2r+1)}= \frac{1}{2},\,
       s_{4\ell-2}^{(2r+1)}=-\frac{1}{2}\Bigr\}                 \nonumber  \\
  &&\hspace{-0.80truecm}\hbox{\lrarrow}\Bigl\{
       s_{4\ell-3}^{(2r)  }=-\frac{1}{2},\,
       s_{4\ell-2}^{(2r)  }= \frac{1}{2},\,
       s_{4\ell-3}^{(2r+1)}=-\frac{1}{2},\,
       s_{4\ell-2}^{(2r+1)}= \frac{1}{2}\Bigr\}\,,                         \\
    \Bigl\{
       s_{4\ell-2}^{(2r-1)}= \frac{1}{2},\,&&\hspace{-0.80truecm}
       S_{4\ell-1}^{(2r-1)},\,
       s_{4\ell-2}^{(2r  )}= \frac{1}{2},\,
       S_{4\ell-1}^{(2r  )}  \Bigr\}                            \nonumber  \\
  &&\hspace{-0.80truecm}\hbox{\lrarrow}\Bigl\{
       s_{4\ell-2}^{(2r-1)}=-\frac{1}{2},\,
       S_{4\ell-1}^{(2r-1)}+1,\,
       s_{4\ell-2}^{(2r  )}=-\frac{1}{2},\,
       S_{4\ell-1}^{(2r  )}+1\Bigr\}\,,                                    \\
    \Bigl\{
       S_{4\ell-1}^{(2r)  },\,&&\hspace{-0.80truecm}
       S_{4\ell  }^{(2r)  },\,
       S_{4\ell-1}^{(2r+1)},\,
       S_{4\ell  }^{(2r+1)}  \Bigr\}                            \nonumber  \\
  &&\hspace{-0.80truecm}\hbox{\lrarrow}\Bigl\{
       S_{4\ell-1}^{(2r)  }+1,\,
       S_{4\ell  }^{(2r)  }-1,\,
       S_{4\ell-1}^{(2r+1)}+1,\,
       S_{4\ell  }^{(2r+1)}-1\Bigr\}\,,                                    \\
    \Bigl\{
       S_{4\ell-1}^{(2r)  }= 1,\,&&\hspace{-0.80truecm}
       S_{4\ell  }^{(2r)  }=-1,\,
       S_{4\ell-1}^{(2r+1)}=1,\,
       S_{4\ell  }^{(2r+1)}=-1  \Bigr\}                         \nonumber  \\
  &&\hspace{-0.80truecm}\hbox{\lrarrow}\Bigl\{
       S_{4\ell-1}^{(2r)  }=-1,\,
       S_{4\ell  }^{(2r)  }= 1,\,
       S_{4\ell-1}^{(2r+1)}=-1,\,
       S_{4\ell  }^{(2r+1)}= 1\Bigr\}\,,                                   \\
    \Bigl\{
       S_{4\ell  }^{(2r-1)},\,&&\hspace{-0.80truecm}
       s_{4\ell+1}^{(2r-1)}=-\frac{1}{2},\,
       S_{4\ell  }^{(2r  )},\,
       s_{4\ell+1}^{(2r  )}=-\frac{1}{2}  \Bigr\}              \nonumber  \\
  &&\hspace{-0.80truecm}\hbox{\lrarrow}\Bigl\{
       S_{4\ell  }^{(2r-1)}-1,\,
       s_{4\ell+1}^{(2r-1)}=+\frac{1}{2},\,
       S_{4\ell  }^{(2r  )}-1,\,
       s_{4\ell+1}^{(2r  )}=+\frac{1}{2}\Bigr\}\,,
\end{subeqnarray}
\end{full}

\noindent
$S_{4\ell-1}^{(r')}$ $(r'\!=\!2r\!-\!1$, $2r$, $2r\!+\!1)$ in eqs.$\,$(3.5b)
and (3.5c) being equal to $-1$ or $0$, and $S_{4\ell}^{(r')}$
$(r'\!=\!2r\!-\!1$, $2r$, $2r\!+\!1)$ in eqs.$\,$(3.5c) and (3.5e) is equal to
$0$ or $1$.

We also carry out the global flips along
the Trotter direction, which are given by~\cite{rf:34}

\begin{full}
\begin{subeqnarray}
\label{eq:3.6}
    \Bigl\{
       s_{\ell'}^{( 1)}=\frac{1}{2},\,&&\hspace{-0.80truecm}
       s_{\ell'}^{( 2)}=\frac{1}{2},\,\cdots,\,
       s_{\ell'}^{(2n)}=\frac{1}{2}\Bigr\}                      \nonumber  \\
  &&\hspace{-0.80truecm}\hbox{\lrarrow}\Bigl\{
       s_{\ell'}^{( 1)}=-\frac{1}{2},\,
       s_{\ell'}^{( 2)}=-\frac{1}{2},\,\cdots,\,
       s_{\ell'}^{(2n)}=-\frac{1}{2}\Bigr\}
                                \qquad(\ell\,'=4\ell-3,\ 4\ell-2)\,,       \\
    \Bigl\{
       S_{\ell'}^{( 1)},\,
       S_{\ell'}^{( 2)},\,&&\hspace{-0.80truecm}\cdots,\,
       S_{\ell'}^{(2n)}\Bigr\}                                  \nonumber  \\
  &&\hspace{-0.80truecm}\hbox{\lrarrow}\Bigl\{
       S_{\ell'}^{( 1)}-1,\,
       S_{\ell'}^{( 2)}-1,\,\cdots,\,
       S_{\ell'}^{(2n)}-1\Bigr\}
                                \qquad(\ell\,'=4\ell-1,\ 4\ell)\,,
\end{subeqnarray}
\end{full}

\noindent
$S_{\ell'}^{(r')}$ $(r'\!=\!1$, $2$, $\cdots$, $2n)$ in eq.$\,$(3.6b) being
equal to $0$ or $1$.  These global flips do change the magnetization in
contrast to the above local flips.  It should be noted that we do not take the
global flips along the real-space direction into account in the present QMC
calculation, assuming that their effect is not serious at least when the
number $N$ of spins is not too small.  This means that we neglect the change
of the winding number of the spin configuration.~\cite{rf:33}

Using this approach, we have investigated numerically the ground-state and
thermodynamic properties of the spin-1/2$\,$-1/2$\,$-1$\,$-1 chain,
confining ourselves to the case where $J_1\!=\!J_3\!=\!1.0$ and
\hbox{$J_2\!>\!0$}.  (Note that $J_1$, or equivalently $J_3$, is chosen to
be the unit of energy.)

\subsection{Ground-state properties}

In this subsection, we discuss the energy gap and the nearest-neighbor
two-spin correlation functions in the ground state, in order to investigate
the ground-state phase transition.

For a finite-$N$ system, we have carried out the QMC calculation only with the
local flips.  Thus, the system is kept in the subspace of a given value of
the $z$-component ${\cal S}_{\rm tot}^z$ of ${\veccalS}_{\rm tot}$.  We have
estimated the energies $E_{{\cal S}_{\rm tot}^z}(N)$ with
\hbox{${\cal S}_{\rm tot}^z\!=\!0$} (the ground state) and $1$ (the
first-excited state), which are expressed as
\begin{equation}
 E_{{\cal S}_{\rm tot}^z}(N)
     = - \bigl\langle Q_1 \bigr\rangle_{{\rm MC},{\cal S}_{\rm tot}^z}\,,
\end{equation}
where

\begin{full}
\begin{eqnarray}
  Q_1 = \sum_{r=1}^n \sum_{\ell=1}^{N/4}\Biggl\{&&
     \biggl(\frac{\partial\rho_{4\ell-3,4\ell-2}^{(2r-1,2r  )}}{\partial\beta}
                 \biggr)\bigg/\rho_{4\ell-3,4\ell-2}^{(2r-1,2r  )}
   + \biggl(\frac{\partial\rho_{4\ell-1,4\ell  }^{(2r-1,2r  )}}{\partial\beta}
                 \biggr)  \bigg/\rho_{4\ell-1,4\ell  }^{(2r-1,2r  )}
                                                             \nonumber    \\
 &&+ \biggl(\frac{\partial\rho_{4\ell-2,4\ell-1}^{(2r  ,2r+1)}}{\partial\beta}
                 \biggr)  \bigg/\rho_{4\ell-2,4\ell-1}^{(2r  ,2r+1)}
   + \Bigl(\frac{\partial\rho_{4\ell  ,4\ell+1}^{(2r  ,2r+1)}}{\partial\beta}
                 \biggr)  \bigg/\rho_{4\ell  ,4\ell+1}^{(2r  ,2r+1)}\Biggl\}
\end{eqnarray}
\end{full}

\noindent
and $\langle\cdots\rangle_{{\rm MC},{\cal S}_{\rm tot}^z}$ stands for the
Monte Carlo average within the subspace determined by the value of
${\cal S}_{\rm tot}^z$.  The energy gap $\Delta(N)$ is defined by
\begin{equation}
\Delta(N) = E_1(N) - E_0(N)\,.
\end{equation}
Carrying out the above-mentioned QMC calculation within the
\hbox{${\cal S}_{\rm tot}^z\!=\!0$} subspace, we have also estimated the three
kinds of ground-state nearest-neighbor two-spin correlation functions
\begin{subeqnarray}
\omega_{1,2}(N) = \langle \vecs_1\cdot\vecs_2\rangle_{{\rm MC},0}
                = 3 \langle s_1^z s_2^z \rangle_{{\rm MC},0}\,,          \\
\omega_{2,3}(N) = \langle \vecs_2\cdot\vecS_3\rangle_{{\rm MC},0}
                = 3 \langle s_2^z S_2^z \rangle_{{\rm MC},0}\,,          \\
\omega_{3,4}(N) = \langle \vecS_3\cdot\vecS_4\rangle_{{\rm MC},0}
                = 3 \langle S_3^z S_4^z \rangle_{{\rm MC},0}\,.
\end{subeqnarray}
In these calculations, the values of $J_2$ have been chosen to be
$J_2\!=1.00$, $0.90$, $0.80$, $0.78$, $0.75$, $0.70$, $0.60$, and $0.50$, and
those of $N$ to be $N\!=\!8$, $16$, $32$, $64$, and $128$.  We have performed
$10^6$ Monte Carlo steps after $10^5$ initial steps for obtaining the thermal
equilibrium.  The Trotter numbers which we have used are $n\!=\!12$, $16$,
$24$, $32$, $40$, and $48$, and the $n$-dependence of the finite-$N$ QMC
results $A_n(N)$ for a given physical quantity $A$ has been extrapolated to
\hbox{$n\!\to\!\infty$} by making a least-squares fit to the
formula~\cite{rf:35}
\begin{equation}
A_n(N) = A(N) + \frac{a}{n^2} + \frac{b}{n^4}\,,
\end{equation}
where $a$ and $b$ are constants which are independent of $n$; $A(N)$ yields
the Trotter-extrapolated value.  Most of the calculations have been done at
the temperature $k_{\rm B}T\!=\!0.05$ only.  For the cases of
\hbox{$J_2\!=\!1.00$} and $0.50$ with $N\!=\!8$ and $16$, however, we have
also done the calculation at \hbox{$k_{\rm B}T\!=\!0.02$}, and have obtained,
for both temperatures, the same Trotter-extrapolated values for $E_0(N)$ and
$E_1(N)$.  These values agree with those obtained by the exact-diagonalization
calculation within the numerical error, although the conservation of the
winding number causes a systematic deviation in short chains.  We therefore
consider that \hbox{$k_{\rm B}T\!=\!0.05$} is low enough to discuss the
zero-temperature properties.  Thus, the calculated result for $E_0(N)$ yields
the ground-state energy $E_{\rm g}(N)$ and those for $\omega_{1,2}(N)$,
$\omega_{2,3}(N)$, and $\omega_{3,4}(N)$ yield the nearest-neighbor two-spin
correlation function in the ground state.

The Trotter-extrapolated values depend upon which data
we use when we make a least-squares fit to the formula given by
eq.$\,$(3.11).  Following Miyashita and Yamamoto's procedure,~\cite{rf:36} we
have performed three different extrapolations.  The first one, denoted by
TEX6, uses six data for $n\!=\!12$, $16$, $24$, $32$, $40$, and $48$, the
second one, TEX5, uses five data for $n\!=\!16$, $24$, $32$, $40$, and $48$,
and the third one, TEX4, uses four data for $n\!=\!24$, $32$, $40$, and
$48$.  Both for the ground-state energy
\hbox{$\epsilon_{\rm g}(N)\!\equiv\!E_0(N)/(N/4)$} per unit cell consisting of
two \hbox{$S\!=\!1/2$} and two \hbox{$S\!=\!1$} spins and for the energy gap
$\Delta(N)$, the three extrapolations give almost the same values down to the
second decimal place.  We determine the Trotter-extrapolated value to be the
average of the three values obtained by the three extrapolations, and estimate
the error by the difference between the average and the farthest value among
the three values.  The results for $\epsilon_{\rm g}(N)$ and $\Delta(N)$ are
tabulated, respectively, in Tables~\ref{table:2} and~\ref{table:3}, where, for
the sake of comparison, the results of the exact-diagonalization calculation
for \hbox{$N\!=\!8$} and $16$ are also listed.

\begin{fulltable}
\caption{Numerical results for the ground-state energy $\epsilon_{\rm g}(N)$
per unit cell in the case where $J_1\!=\!J_3\!=\!1.0$.  The values in the
columns denoted by TEX give the Totter-extrapolated results, and those in the
columns denoted by ED give the results of the exact-diagonalization
calculation.  The extrapolated values $\epsilon_{\rm g}(\infty)$ to
$N\!\to\!\infty$ are given in the rightmost ($N\!\to\!\infty$) column.  It is
noted that the figures in the parentheses show errors in the last digit.
}
\label{table:2}
\begin{fulltabular}{@{\hspace{\tabcolsep}\extracolsep{\fill}}ccccccccc}
  \hline
%
  & \multicolumn{2}{c}{$N\!=\!8$} & \multicolumn{2}{c}{$N\!=\!16$} & $N\!=\!32$
  & $N\!=\!64$ & $N\!=\!128$ & $N\!\to\!\infty$
  \\
%
%
  & TEX & ED & TEX & ED & TEX & TEX & TEX &                   \\
  \hline
 $J_2\!=\!1.00$ & $-3.54\,(1)$ & $-3.5470$    & $-3.52\,(1)$ & $-3.5159$
                & $-3.52\,(1)$ & $-3.52\,(1)$ & $-3.52\,(1)$ & $-3.52\,(1)$ \\
 $J_2\!=\!0.90$ & $-3.37\,(1)$ & $-3.3832$    & $-3.34\,(1)$ & $-3.3449$
                & $-3.34\,(1)$ & $-3.34\,(1)$ & $-3.34\,(1)$ & $-3.34\,(1)$ \\
 $J_2\!=\!0.80$ & $-3.22\,(1)$ & $-3.2326$    & $-3.19\,(1)$ & $-3.1891$
                & $-3.18\,(1)$ & $-3.18\,(1)$ & $-3.18\,(1)$ & $-3.18\,(1)$ \\
 $J_2\!=\!0.78$ & $-3.19\,(1)$ & $-3.2045$    & $-3.16\,(1)$ & $-3.1609$
                & $-3.15\,(1)$ & $-3.15\,(1)$ & $-3.15\,(1)$ & $-3.15\,(1)$ \\
 $J_2\!=\!0.75$ & $-3.15\,(1)$ & $-3.1638$    & $-3.12\,(1)$ & $-3.1209$
                & $-3.11\,(1)$ & $-3.11\,(1)$ & $-3.11\,(1)$ & $-3.11\,(1)$ \\
 $J_2\!=\!0.70$ & $-3.09\,(1)$ & $-3.1002$    & $-3.06\,(1)$ & $-3.0611$
                & $-3.05\,(1)$ & $-3.05\,(1)$ & $-3.05\,(1)$ & $-3.05\,(1)$ \\
 $J_2\!=\!0.60$ & $-2.98\,(1)$ & $-2.9908$    & $-2.97\,(1)$ & $-2.9655$
                & $-2.96\,(1)$ & $-2.96\,(1)$ & $-2.96\,(1)$ & $-2.96\,(1)$ \\
 $J_2\!=\!0.50$ & $-2.90\,(1)$ & $-2.9060$    & $-2.89\,(1)$ & $-2.8937$
                & $-2.89\,(1)$ & $-2.89\,(1)$ & $-2.89\,(1)$ & $-2.89\,(1)$ \\
\hline
\end{fulltabular}
\end{fulltable}

\begin{fulltable}
\caption{Numerical results for the energy gap $\Delta(N)$ in the case where
$J_1\!=\!J_3\!=\!1.0$.  The values in the columns denoted by TEX give the
Totter-extrapolated results, and those in the columns denoted by ED give the
results of the exact-diagonalization calculation.  The extrapolated values
$\Delta(\infty)$ to $N\!\to\!\infty$ are given in the rightmost
($N\!\to\!\infty$) column.  It is noted that the figures in the parentheses
show errors in the last digit.
}
\label{table:3}
\begin{fulltabular}{@{\hspace{\tabcolsep}\extracolsep{\fill}}ccccccccc}
  \hline
%
  & \multicolumn{2}{c}{$N\!=\!8$} & \multicolumn{2}{c}{$N\!=\!16$} & $N\!=\!32$
  & $N\!=\!64$ & $N\!=\!128$ & $N\!\to\!\infty$
  \\
%
%
  & TEX & ED & TEX & ED & TEX & TEX & TEX &                   \\
  \hline
 $J_2\!=\!1.00$ & $0.61\,(1)$ & $0.6156$    & $0.46\,(1)$ & $0.4497$
                & $0.44\,(1)$ & $0.44\,(1)$ & $0.44\,(1)$ & $0.44\,(1)$   \\
 $J_2\!=\!0.90$ & $0.55\,(1)$ & $0.5566$    & $0.36\,(1)$ & $0.3522$
                & $0.32\,(1)$ & $0.31\,(1)$ & $0.31\,(1)$ & $0.31\,(1)$   \\
 $J_2\!=\!0.80$ & $0.51\,(1)$ & $0.5126$    & $0.28\,(1)$ & $0.2807$
                & $0.17\,(1)$ & $0.13\,(1)$ & $0.13\,(1)$ & $0.13\,(1)$   \\
 $J_2\!=\!0.78$ & $0.50\,(1)$ & $0.5064$    & $0.27\,(1)$ & $0.2733$
                & $0.15\,(1)$ & $0.09\,(1)$ & $0.06\,(2)$ & $0.06\,(2)$   \\
 $J_2\!=\!0.75$ & $0.50\,(1)$ & $0.4992$    & $0.27\,(1)$ & $0.2685$
                & $0.14\,(1)$ & $0.08\,(1)$ & $0.06\,(2)$ & $0.06\,(2)$   \\
 $J_2\!=\!0.70$ & $0.49\,(1)$ & $0.4930$    & $0.28\,(1)$ & $0.2785$
                & $0.20\,(1)$ & $0.18\,(1)$ & $0.18\,(1)$ & $0.18\,(1)$   \\
 $J_2\!=\!0.60$ & $0.51\,(1)$ & $0.5059$    & $0.37\,(1)$ & $0.3555$
                & $0.34\,(1)$ & $0.34\,(1)$ & $0.34\,(1)$ & $0.34\,(1)$   \\
 $J_2\!=\!0.50$ & $0.56\,(1)$ & $0.5529$    & $0.48\,(1)$ & $0.4697$
                & $0.48\,(1)$ & $0.48\,(1)$ & $0.48\,(1)$ & $0.48\,(1)$   \\
\hline
\end{fulltabular}
\end{fulltable}

In order to obtain the results in the thermodynamic limit
(\hbox{$N\!\to\!\infty$}), the Trotter-extrapolated values $A(N)$ have further
been extrapolated by the least-squares method using a linear function of
$1/N^2$,
\begin{equation}
A(N) = A(\infty) + \frac{c}{N^2}
\end{equation}
with a constant $c$.  Performing the extrapolation, we have used three
data $A(32)$, $A(64)$, and $A(128)$ for $J_2\!=0.75$ and $0.78$, and four
data $A(16)$, $A(32)$, $A(64)$, and $A(128)$ for the other values of
$J_2$.  Figure~\ref{fig:4} illustrate this extrapolation for the energy gap,
where $A$ is $\Delta$.  The $N\!\to\!\infty$ extrapolated values
$\epsilon_{\rm g}(\infty)$ for the ground-state energy per unit cell and
those $\Delta(\infty)$ for the energy gap are also listed in
Tables~\ref{table:2} and~\ref{table:3}, respectively, and are plotted as a
function of $J_2$ in Fig.~\ref{fig:5} and Fig.~\ref{fig:6},
respectively.  From Fig.~\ref{fig:6} we see that $\Delta(\infty)$ vanishes
at
\begin{equation}
  J_2 = J_{2{\rm c}} = 0.77\pm0.01\,.
\end{equation}
This result shows that the ground state of the present system with
$J_1\!=\!J_3\!=\!1.0$ undergoes a second-order phase transition at this value
of $J_2$.

The Trotter extrapolated values $\omega_{1,2}(N)$,
$\omega_{2,3}(N)$, and $\omega_{3,4}(N)$ for the nearest-neighbor
two-spin correlation functions are tabulated in
Tables~\ref{table:4},~\ref{table:5}, and~\ref{table:6}, respectively,
where the corresponding $N\!\to\!\infty$ extrapolated values,
$\omega_{1,2}(\infty)$, $\omega_{2,3}(\infty)$, and
$\omega_{3,4}(\infty)$ are also listed.  In Fig.~\ref{fig:7} we plot
$\omega_{1,2}(\infty)$, $\omega_{2,3}(\infty)$, and $\omega_{3,4}(\infty)$ as
functions of $J_2$, where each correlation function is normalized by its
minimum value $\bigl($note that \hbox{$-3/4\!\le\!\omega_{1,2}(N)\!\le\!1/4$},
\hbox{$-1\!\le\!\omega_{2,3}(N)\!\le\!1/2$}, and
\hbox{$-2\!\le\!\omega_{3,4}(N)\!\le\!1$}$\bigr)$.  This figure demonstrates
that as $J_2$ increases, $\vert\omega_{1,2}(\infty)\vert$ and
$\vert\omega_{2,3}(\infty)\vert$ rapidly decreases and increases,
respectively, around \hbox{$J_2\!\sim\!J_{2{\rm c}}$}, while
$\vert\omega_{3,4}(\infty)\vert$ rather gradually decreases with increasing
$J_2$.  Giving a thought to these $J_2$-dependences of the correlation
functions, we may schematically represent, by means of the VBS
picture,~\cite{rf:20} the ground states for \hbox{$J_2\!<\!J_{2{\rm c}}$} and
that for \hbox{$J_2\!>\!J_{2{\rm c}}$} as depicted in Fig.~\ref{fig:8}.  The
VBS wave function of Fig.~\ref{fig:8}(a) corresponds to the state for
\hbox{$J_2\!=\!0.0$}, while the VBS wave function of Fig.~\ref{fig:8}(b) to
the state for \hbox{$J_1\!=\!0.0$} and \hbox{$J_2\!\to\!+\infty$}.  The change
of the correlation functions as a function of $J_2$ clearly illustrates the
interpolation between these two limiting states.

\begin{fulltable}
\caption{Trotter-extrapolated ground-state nearest-neighbor two-spin
correlation function $\omega_{1,2}(N)$ and the \hbox{$N\!\to\!\infty$}
extrapolated ground-state nearest-neighbor two-spin correlation function
$\omega_{1,2}(\infty)$ in the case where $J_1\!=\!J_3\!=\!1.0$.  It is noted
that the figures in the parentheses show errors in the last digit.
}
\label{table:4}
\begin{fulltabular}{@{\hspace{\tabcolsep}\extracolsep{\fill}}ccccccc}
  \hline
  & $N\!=\!8$ & $N\!=\!16$ & $N\!=\!32$ & $N\!=\!64$ & $N\!=\!128$ 
  & $N\!\to\!\infty$
  \\
  \hline
 $J_2\!=\!1.00$ & $-0.28\,(1)$ & $-0.23\,(1)$ & $-0.22\,(1)$
                & $-0.22\,(1)$ & $-0.22\,(1)$ & $-0.22\,(1)$     \\
 $J_2\!=\!0.90$ & $-0.33\,(1)$ & $-0.27\,(1)$ & $-0.26\,(1)$
                & $-0.26\,(1)$ & $-0.26\,(1)$ & $-0.26\,(1)$     \\
 $J_2\!=\!0.80$ & $-0.41\,(1)$ & $-0.37\,(1)$ & $-0.35\,(1)$
                & $-0.34\,(1)$ & $-0.34\,(1)$ & $-0.34\,(1)$     \\
 $J_2\!=\!0.78$ & $-0.43\,(1)$ & $-0.40\,(1)$ & $-0.39\,(1)$
                & $-0.38\,(1)$ & $-0.38\,(1)$ & $-0.38\,(1)$     \\
 $J_2\!=\!0.75$ & $-0.46\,(1)$ & $-0.46\,(1)$ & $-0.46\,(1)$
                & $-0.46\,(1)$ & $-0.46\,(1)$ & $-0.46\,(1)$     \\
 $J_2\!=\!0.70$ & $-0.51\,(1)$ & $-0.53\,(1)$ & $-0.54\,(1)$
                & $-0.54\,(1)$ & $-0.54\,(1)$ & $-0.54\,(1)$     \\
 $J_2\!=\!0.60$ & $-0.60\,(1)$ & $-0.62\,(1)$ & $-0.62\,(1)$
                & $-0.62\,(1)$ & $-0.62\,(1)$ & $-0.62\,(1)$     \\
 $J_2\!=\!0.50$ & $-0.66\,(1)$ & $-0.67\,(1)$ & $-0.67\,(1)$
                & $-0.67\,(1)$ & $-0.67\,(1)$ & $-0.67\,(1)$     \\
\hline
\end{fulltabular}
\end{fulltable}

\begin{fulltable}
\caption{Trotter-extrapolated ground-state nearest-neighbor two-spin
correlation function $\omega_{2,3}(N)$ and the \hbox{$N\!\to\!\infty$}
extrapolated ground-state nearest-neighbor two-spin correlation function
$\omega_{2,3}(\infty)$ in the case where $J_1\!=\!J_3\!=\!1.0$.  It is noted
that the figures in the parentheses show errors in the last digit.
}
\label{table:5}
\begin{fulltabular}{@{\hspace{\tabcolsep}\extracolsep{\fill}}ccccccc}
  \hline
  & $N\!=\!8$ & $N\!=\!16$ & $N\!=\!32$ & $N\!=\!64$ & $N\!=\!128$ 
  & $N\!\to\!\infty$
  \\
  \hline
 $J_2\!=\!1.00$ & $-0.87\,(1)$ & $-0.89\,(1)$ & $-0.89\,(1)$
                & $-0.89\,(1)$ & $-0.89\,(1)$ & $-0.89\,(1)$     \\
 $J_2\!=\!0.90$ & $-0.81\,(1)$ & $-0.84\,(1)$ & $-0.85\,(1)$
                & $-0.85\,(1)$ & $-0.85\,(1)$ & $-0.85\,(1)$     \\
 $J_2\!=\!0.80$ & $-0.72\,(1)$ & $-0.74\,(1)$ & $-0.75\,(1)$
                & $-0.76\,(1)$ & $-0.76\,(1)$ & $-0.76\,(1)$     \\
 $J_2\!=\!0.78$ & $-0.70\,(1)$ & $-0.70\,(1)$ & $-0.71\,(1)$
                & $-0.72\,(1)$ & $-0.72\,(1)$ & $-0.72\,(1)$     \\
 $J_2\!=\!0.75$ & $-0.67\,(1)$ & $-0.64\,(1)$ & $-0.63\,(1)$
                & $-0.62\,(1)$ & $-0.62\,(1)$ & $-0.62\,(1)$     \\
 $J_2\!=\!0.70$ & $-0.61\,(1)$ & $-0.54\,(1)$ & $-0.52\,(1)$
                & $-0.51\,(1)$ & $-0.51\,(1)$ & $-0.51\,(1)$     \\
 $J_2\!=\!0.60$ & $-0.47\,(1)$ & $-0.41\,(1)$ & $-0.40\,(1)$
                & $-0.40\,(1)$ & $-0.40\,(1)$ & $-0.40\,(1)$     \\
 $J_2\!=\!0.50$ & $-0.35\,(1)$ & $-0.31\,(1)$ & $-0.31\,(1)$
                & $-0.31\,(1)$ & $-0.31\,(1)$ & $-0.31\,(1)$     \\
\hline
\end{fulltabular}
\end{fulltable}

\begin{fulltable}
\caption{Trotter-extrapolated ground-state nearest-neighbor two-spin
correlation function $\omega_{3,4}(N)$ and the \hbox{$N\!\to\!\infty$}
extrapolated ground-state nearest-neighbor two-spin correlation function
$\omega_{3,4}(\infty)$ in the case where $J_1\!=\!J_3\!=\!1.0$.  It is noted
that the figures in the parentheses show errors in the last digit.
}
\label{table:6}
\begin{fulltabular}{@{\hspace{\tabcolsep}\extracolsep{\fill}}ccccccc}
 \hline
  & $N\!=\!8$ & $N\!=\!16$ & $N\!=\!32$ & $N\!=\!64$ & $N\!=\!128$ 
  & $N\!\to\!\infty$
  \\
 \hline
 $J_2\!=\!1.00$ & $-1.58\,(1)$ & $-1.53\,(1)$ & $-1.52\,(1)$
                & $-1.52\,(1)$ & $-1.52\,(1)$ & $-1.52\,(1)$     \\
 $J_2\!=\!0.90$ & $-1.62\,(1)$ & $-1.57\,(1)$ & $-1.56\,(1)$
                & $-1.56\,(1)$ & $-1.56\,(1)$ & $-1.56\,(1)$     \\
 $J_2\!=\!0.80$ & $-1.69\,(1)$ & $-1.66\,(1)$ & $-1.64\,(1)$
                & $-1.63\,(1)$ & $-1.63\,(1)$ & $-1.63\,(1)$     \\
 $J_2\!=\!0.78$ & $-1.70\,(1)$ & $-1.68\,(1)$ & $-1.67\,(1)$
                & $-1.66\,(1)$ & $-1.66\,(1)$ & $-1.66\,(1)$     \\
 $J_2\!=\!0.75$ & $-1.73\,(1)$ & $-1.73\,(1)$ & $-1.72\,(1)$
                & $-1.72\,(1)$ & $-1.72\,(1)$ & $-1.72\,(1)$     \\
 $J_2\!=\!0.70$ & $-1.78\,(1)$ & $-1.79\,(1)$ & $-1.80\,(1)$
                & $-1.80\,(1)$ & $-1.80\,(1)$ & $-1.80\,(1)$     \\
 $J_2\!=\!0.60$ & $-1.86\,(1)$ & $-1.87\,(1)$ & $-1.87\,(1)$
                & $-1.87\,(1)$ & $-1.87\,(1)$ & $-1.87\,(1)$     \\
 $J_2\!=\!0.50$ & $-1.92\,(1)$ & $-1.92\,(1)$ & $-1.92\,(1)$
                & $-1.92\,(1)$ & $-1.92\,(1)$ & $-1.92\,(1)$     \\
 \hline
\end{fulltabular}
\end{fulltable}

\subsection{Thermodynamic properties}

We now turn to the discussion of the thermodynamic properties.  Performing
the QMC calculation which takes into account the global flips given by
eqs.$\,$(3.5a) and (3.5b) as well as the local flips given by
eqs.$\,$(3.5a)-(3.5e), we have calculated the temperature dependences of
the specific heat $C(N)$ and the magnetic susceptibility $\chi(N)$ for
\hbox{$J_2\!=\!J_{2{\rm c}}\!=\!0.77$}, and also for \hbox{$J_2\!=\!1.0$}
and $5.0$.  Here, the expressions for $C(N)$~\cite{rf:37} and $\chi(N)$ are
given by
\begin{equation}
 C(N) = \frac{1}{k_{\rm B}T^2}
           \Big( \bigl\langle Q_1^2 \bigr\rangle_{{\rm MC}}
               - \bigl\langle Q_1   \bigr\rangle_{{\rm MC}}^2
               + \bigl\langle Q_2   \bigr\rangle_{{\rm MC}} \Big)
\end{equation}
and
\begin{equation}
 \chi(N) = \frac{1}{T} 
    \Bigl\langle \bigl({\cal S}_{\rm tot}^z\bigr)^2 \Bigr\rangle_{{\rm MC}}\,,
\end{equation}
where $\langle\cdots\rangle_{{\rm MC}}$ denotes the Monte Carlo average, $Q_1$
is given by eq.$\,$(3.8), and

\begin{full}
\begin{eqnarray}
 Q_2 = \sum_{r=1}^n \sum_{\ell=1}^{N/4}\Biggl[&&
     \biggl(\frac{\partial^2\rho_{4\ell-3,4\ell-2}^{(2r-1,2r  )}}
        {\partial\beta^2}\biggr)\bigg/\rho_{4\ell-3,4\ell-2}^{(2r-1,2r  )}
   - \Biggl\{\biggl(\frac{\partial\rho_{4\ell-3,4\ell-2}^{(2r-1,2r  )}}
        {\partial\beta}\biggr)\bigg/\rho_{4\ell-3,4\ell-2}^{(2r-1,2r  )}
     \Biggr\}^2                                              \nonumber    \\
 &&+ \biggl(\frac{\partial^2\rho_{4\ell-1,4\ell  }^{(2r-1,2r  )}}
        {\partial\beta^2}\biggr)\bigg/\rho_{4\ell-1,4\ell  }^{(2r-1,2r  )}
   - \Biggl\{\biggl(\frac{\partial\rho_{4\ell-1,4\ell  }^{(2r-1,2r  )}}
        {\partial\beta}\biggr)\bigg/\rho_{4\ell-1,4\ell  }^{(2r-1,2r  )}
     \Biggr\}^2                                              \nonumber    \\
 &&+ \biggl(\frac{\partial^2\rho_{4\ell-2,4\ell-1}^{(2r  ,2r+1)}}
        {\partial\beta^2}\biggr)  \bigg/\rho_{4\ell-2,4\ell-1}^{(2r  ,2r+1)}
   - \Biggl\{\biggl(\frac{\partial\rho_{4\ell-2,4\ell-1}^{(2r  ,2r+1)}}
        {\partial\beta}\biggr)  \bigg/\rho_{4\ell-2,4\ell-1}^{(2r  ,2r+1)}
     \Biggr\}^2                                              \nonumber    \\
 &&+ \biggl(\frac{\partial^2\rho_{4\ell  ,4\ell+1}^{(2r  ,2r+1)}}
        {\partial\beta^2}\biggr)  \bigg/\rho_{4\ell  ,4\ell+1}^{(2r  ,2r+1)}
   - \Biggl\{\biggl(\frac{\partial\rho_{4\ell  ,4\ell+1}^{(2r  ,2r+1)}}
        {\partial\beta}\biggr)  \bigg/\rho_{4\ell  ,4\ell+1}^{(2r  ,2r+1)}
     \Biggr\}^2\,\Biggr]\,.
\end{eqnarray}
\end{full}

\noindent
In deriving eq.$\,$(3.15), we assume that the g-factor associated with the
$S\!=\!1/2$ spin is identical to that associated with the $S\!=\!1$ spin
and that $\langle{\cal S}_{\rm tot}^z\rangle_{{\rm MC}}$ vanishes.  For
$J_2\!=\!1.0$ and $5.0$, the values of $N$ have been chosen only to be
$N\!=\!8$, $16$ and $32$.  This is because for these $J_2$'s, we expect weak
$N$-dependences of $C(N)/(N/4)$ and $\chi(N)/(N/4)$, since the energy gap
$\Delta(\infty)$ is rather large (see Fig.~\ref{fig:6}), or equivalently, the
correlation length is rather short.  For the critical value $J_2\!=\!0.77$, on
the other hand, the calculation has been done also for $N\!=\!64$ and $128$
when $T$ is not too high, in addition to $N\!=\!8$, $16$ and $32$.  We list
in Table~\ref{table:7} Trotter numbers $n's$, the Monte Carlo steps, and the
initial steps for obtaining the thermal equilibrium.  By using the QMC results
for all of the four Trotter numbers, the $n$-dependence of $C_n(N)$ for
$J_2\!=\!0.77$ and $1.0$ has been extrapolated to $n\!\to\!\infty$ by the
least-squares method with the formula of eq.$\,$(3.11).~\cite{rf:35}  Similar
extrapolations for $C_n(N)$ for $J_2\!=\!5.0$ and for $\chi_n(N)$ for
$J_2\!=\!0.77$, $1.0$, and $5.0$ have been done by the use of the
formula~\cite{rf:35}
\begin{equation}
A_n(N) = A(N) + \frac{a}{n^2}\,,
\end{equation}
where $a$ is a constant, instead of the formula of eq.$\,$(3.11).

\begin{fulltable}
\caption{Sets of the Trotter numbers (n) used at various temperatures, and
also the Monte Carlo steps (MCS) and the initial steps (IS) spent for each
Trotter number at each temperature, where the former steps do not include
the latter steps.  Note that $J_1\!=\!J_3\!=\!1.0$.
}
\label{table:7}
\begin{fulltabular}{@{\hspace{\tabcolsep}\extracolsep{\fill}}ccccc}
 \hline
 $k_{\rm B}T$ for $J_2\!=\!0.77$, $1.0$ & $k_{\rm B}T$ for $J_2\!=\!5.0$ &
 $n$ & MCS & IS                                                           \\
 \hline
%
%
   $0.08 \!\leq\!k_{\rm B}T\!\leq\!0.10 $ &
   $0.400\!\leq\!k_{\rm B}T\!\leq\!0.625$ &
   $12$,  $16$,  $24$,  $32$ &  $120\!\times\!10^4$ & $6\!\times\!10^4$    \\
  $k_{\rm B}T\!=\!0.15 $                  &
  $k_{\rm B}T\!=\!0.750$                  &
  $\ 8$,  $12$,  $16$,  $24$ &  $100\!\times\!10^4$ & $5\!\times\!10^4$    \\
%
                                          &
  $k_{\rm B}T\!=\!0.875$                  &
  $\ 8$,  $12$,  $16$,  $24$ & $\ 50\!\times\!10^4$ & $5\!\times\!10^4$    \\
%
  $k_{\rm B}T\!=\!0.20 $                  &
  $1.000\!\leq\!k_{\rm B}T\!\leq\!1.375$ &
  $\ 6$, $\ 8$,  $12$,  $16$ & $\ 50\!\times\!10^4$ & $5\!\times\!10^4$    \\
  $0.30 \!\leq\!k_{\rm B}T\!\leq\!0.50 $ &
  $1.500\!\leq\!k_{\rm B}T\!\leq\!2.750$ &
  $\ 4$, $\ 6$, $\ 8$,  $12$ & $\ 40\!\times\!10^4$ & $5\!\times\!10^4$    \\
  $0.60 \!\leq\!k_{\rm B}T\!\leq\!1.00 $ &
  $3.000\!\leq\!k_{\rm B}T\!\leq\!5.000$ &
  $\ 2$, $\ 4$, $\ 6$, $\ 8$ & $\ 30\!\times\!10^4$ & $5\!\times\!10^4$    \\
   $1.20 \!\leq\!k_{\rm B}T\!\leq\!2.00 $ &
   $6.000\!\leq\!k_{\rm B}T\!\leq\!8.000$ &
  $\ 2$, $\ 4$, $\ 6$, $\ 8$ & $\ 25\!\times\!10^4$ & $5\!\times\!10^4$    \\
\hline
\end{fulltabular}
\end{fulltable}

In Fig.~\ref{fig:9} we plot versus the temperature $T$ the
Trotter-extrapolated values $C(N)/(N/4)$ $[$Fig.~\ref{fig:9}(a)$]$ with
\hbox{$N\!=\!8$}, $16$, and $32$ for the specific heat per unit cell and
$\chi(N)/(N/4)$ $[$Fig.~\ref{fig:9}(b)$]$ with \hbox{$N\!=\!8$}, $16$,
and $32$ for the magnetic susceptibility per unit cell, which have been
obtained for \hbox{$J_2\!=\!1.0$}.  For the sake of comparison, we also plot,
in these figures, the exact results for both quantities for \hbox{$N\!=\!8$}
and \hbox{$J_2\!=\!1.0$} versus $T$, which we have obtained by diagonalizing
completely the Hamiltonian to calculate all the eigenvalues.  The
corresponding plots for \hbox{$J_2\!=\!0.77$} are presented in
Fig.~\ref{fig:10}.  As is expected, both $C(N)/(N/4)$ and $\chi(N)/(N/4)$ for
\hbox{$J_2\!=\!1.0$} have rather weak $N$-dependences even at
\hbox{$k_{\rm B}T\!=\!0.08$}.  For \hbox{$J_2\!=\!0.77$}, on the other hand,
the $N$-dependence of $\chi(N)/(N/4)$ is noticeable at low temperatures; it
increases as $N$ increases.  Judging from the $T$-dependences of $\chi(64)/16$
and $\chi(128)/32$, we may conclude that at \hbox{$T\!=\!0.0$}, the
magnetic susceptibility per unit cell in the thermodynamic limit is finite for
\hbox{$J_2\!=\!0.77$} with the value
\hbox{[$\chi(N)/(N/4)]_{N\!\to\infty}\!=\!0.42\pm0.01$}.  This is consistent
with the fact that the energy gap vanishes for this value of $J_2$.

Figure~\ref{fig:11} presents the plots versus $T$ of $C(N)/(N/4)$
$[$Fig.~\ref{fig:11}(a)$]$ with \hbox{$N\!=\!8$}, $16$, and $32$ and
$\chi(N)/(N/4)$ $[$Fig.~\ref{fig:11}(b)$]$ with \hbox{$N\!=\!8$}, $16$, and
$32$, which have been obtained for $J_2\!=\!5.0$, and also the plots versus
$T$ of the exact results for both quantities for $N\!=\!8$ and
$J_2\!=\!5.0$.  It is noted that the $N$-dependences of both $C(N)/(N/4)$ and
$\chi(N)/(N/4)$ are weak as in the case of $J_2\!=\!1.0$.  A characteristic
feature of the $C(N)/(N/4)$ versus $T$ curve shown in Fig.~\ref{fig:11}(a) is
that it has a double peak.  The origin of the two peaks can
be understood as follows.  The dot-dashed line in Fig.~\ref{fig:11}(a) shows
the $T$-dependence of the specific heat for two pairs of the two-spin system
consisting of an \hbox{$S\!=\!1/2$} spin $\vecs$ and an \hbox{$S\!=\!1$}
spin $\vecS$ which couple with each other by $5\vecs\cdot\vecS$.  Comparing
this with the $T$-dependence of $C(N)/(N/4)$, we may conclude that the
higher-temperature peak is associated with the Schottky-type peak of this
two-spin system.  On the other hand, the dashed line in Fig.~\ref{fig:11}(a)
shows the $T$-dependence of the half of the specific heat for the $S\!=\!1/2$
bond-alternating chain described by the Hamiltonian ${\cal H}_{(a)}$ $[$see
eq.$\,$(2.3)$]$ with $J_1\!=\!J_3\!=\!1.0$ and $N\!=\!8$.  The peak height
as well as the peak position in this $T$-dependence agrees, respectively, with
those in the $T$-dependence of $C(N)/(N/4)$, which may leads us to the
conclusion that the lower-temperature peak is associated with the peak coming
from the short-range order in the $S\!=\!1/2$ bond-alternating chain.  A
somewhat unusual behavior of $\chi(N)/(N/4)$ as a function of $T$, seen in
Fig.~\ref{fig:11}(b), can also be understood as follows.  At high temperatures
the $T$-dependence of $\chi(N)/(N/4)$ should be similar to that of the
magnetic susceptibility for two pairs of the above two-spin system, which
diverges at \hbox{$T\!\to\!0.0$} $[$see the dot-dashed line in
Fig.~\ref{fig:11}(b)$]$, while $\chi(N)/(N/4)$ should vanish at
\hbox{$T\!=\!0.0$} since the present system has massive excitations when
$J_2\!=\!5.0$.

\section{Summary and Conclusions}

We have explored the ground-state properties of the quantum mixed spin system
described by the Hamiltonian ${\cal H}$
$\bigl[$see eq.$\,$(1.1)$\bigr]$.  First, we have applied the Lieb-Mattis
theorem~\cite{rf:17} to show that the ground state of the system is
nonmagnetic when \hbox{$J_1\!>\!0$} and when \hbox{$J_1\!<\!0$} and
\hbox{$J_3\!>\!0$}; in the remaining region where \hbox{$J_1\!<\!0$} and
\hbox{$J_3\!<\!0$} it is ferrimagnetic or ferromagnetic depending upon
whether \hbox{$J_2\!>\!0$} or \hbox{$J_2\!<\!0$}
$\bigl[$Table~\ref{table:1}$\,\bigr]$.  Second, applying the Wigner-Eckart
theorem~\cite{rf:19} to the limiting cases (a)
\hbox{$\vert J_2\vert\!\gg\!\vert J_1\vert,\;\vert J_3\vert$}
with \hbox{$J_2\!>\!0$},
(b) \hbox{$\vert J_2\vert\!\gg\!\vert J_1\vert,\;\vert J_3\vert$} with
\hbox{$J_2\!<\!0$}, and (c)
\hbox{$\vert J_1\vert\!\gg\!\vert J_2\vert,\;\vert J_3\vert$}
with \hbox{$J_1\!<\!0$}, we have mapped the Hamiltonian ${\cal H}$ into the
bond-alternating chains described by the Hamiltonians ${\cal H}_{(a)}$ with
effective \hbox{$S\!=\!1/2$} spins $\bigl[$eq.$\,$(2.3)$\bigr]$,
${\cal H}_{(b)}$ with effective \hbox{$S\!=\!3/2$} spins
$\bigl[$eq.$\,$(2.5)$\bigr]$, and ${\cal H}_{(c)}$ with effective
\hbox{$S\!=\!1$} spins $\bigl[$eq.$\,$(2.7)$\bigr]$, respectively.  This leads
to the facts that in the limiting case (a) the present system is massless when
eq.$\,$(2.8) holds, and in the limiting case (b) it is massless when
eq.$\,$(2.9) holds.
We have also carried out the perturbation calculations to show that in the
limiting cases (d)
\hbox{$J_3\!\gg\!\vert J_1\vert,\;\vert J_2\vert$} and (e)
\hbox{$J_1\!\gg\!\vert J_2\vert$} and \hbox{$J_1\!\gg\!J_3\!>\!0$}, the
system is massless when eqs.$\,$(2.14) and (2.18) hold, respectively.
These results for \hbox{$J_3\!=\!1.0$} are summarized in
Fig.~\ref{fig:2}.  Third, performing the QMC calculation without the global
flips at a sufficiently low temperature (\hbox{$k_{\rm B}T\!=\!0.05$}), we
have shown that when \hbox{$J_1\!=\!J_3\!=\!1.0$}, the energy gap
$\Delta(\infty)$ vanishes at \hbox{$J_2\!=\!J_{2{\rm c}}\!=\!0.77\pm0.01$}
$\bigl[$Fig.~\ref{fig:6}$\bigr]$, where the second-order phase transition
occurs in the ground state.  The representation of both the ground state for
$J_2\!<\!J_{2{\rm c}}$ and that for $J_2\!>\!J_{2{\rm c}}$ by means of the VBS
picture is given in Fig.~\ref{fig:8}, which is suggested from the $J_2$
dependences $\bigl[$Fig.~\ref{fig:7}$\bigr]$ of the ground-state
nearest-neighbor two-spin correlation functions $\omega_{1,2}(\infty)$,
$\omega_{2,3}(\infty)$, and $\omega_{3,4}(\infty)$.

Furthermore, performing the QMC calculation including the global flips along
the Trotter direction as well as the local flips, we have calculated the
temperature dependences of the specific heat $C(N)/(N/4)$ per unit cell and
the magnetic susceptibility $\chi(N)/(N/4)$ per unit cell with $N\!=\!8$,
$16$, and $32$ for \hbox{$J_2\!=\!0.77$}, $1.0$, and $5.0$ with $J_1$ and
$J_3$ fixed at \hbox{$J_1\!=\!J_3\!=\!1.0$}.  The results of these
calculations are depicted in Figs.$\,$9-11.

Figure~2 clearly demonstrates that several massless lines exist on the
$J_1$ versus $J_2$ plane with \hbox{$J_1\!>\!0$} in the case of
\hbox{$J_3\!=\!1.0$}.  We are now exploring the details of the ground-state
phase diagram on this plane by the use of a density-matrix
renormalization-group method proposed originally by White.~\cite{rf:38}  Our
preliminary results show that there exist four massless lines which divides
the upper-half plane into six regions; the ground state in each region can be
understood by the VBS picture.~\cite{rf:31}

In conclusion, we hope that the present study stimulates further experimental
studies on related subject, which include the synthesization of quantum
mixed spin systems with nonmagnetic ground states.

\section*{Acknowledgements}
We would like to thank Professor S.~Yamamoto, Dr.~T.~Fukui, and Dr.~K.~Totsuka
for invaluable discussions.  We also thank the Supercomputer Center, Institute
for Solid State Physics, University of Tokyo and the Computer Center, Tohoku
University for computational facilities.  One of us (H.-J.~M.) gratefully
acknowledges the financial support from the Japan Society for the Promotion
of Science.  The present work has been supported in part by a Grant-in-Aid for
Scientific Research (C) and for International Scientific Research (Joint
Research) from the Ministry of Education, Science, Sports and Culture.

\newpage

\begin{figure}
\caption{Unit cells of the lowest-energy spin arrays, for (a) $J_3\!>\!0$
and (b) $J_3\!<\!0$, of the corresponding classical system in which
$\vecs_\ell$ and $\vecS_\ell$ are replaced, respectively, by the classical
spin vectors with the magnitudes of $1/2$ and $1$.  The short and long arrows
indicate the former and latter classical spin vectors, respectively.}
\label{fig:1}
\end{figure}

\begin{figure}
\caption{Massless lines in the $J_1$ versus $J_2$ plane with $J_1\!>\!0$ and
$J_3\!=\!1.0$ in the limiting cases (a), (b), (d), and (e).  The massless
lines in the cases (a), (b), and (e) are shown by the dashes lines, while the
massless line in the case (d) is by the solid line in the box where the scales
of both the abscissa and the ordinate are enlarged by factor $10$.  The
dotted lines show simple extrapolations of the solid line.  The cirle denotes
the massless point ($J_2\!=\!0.77$, $J_1\!=\!1.0$), which is obtained in
\S3.2 $[$see eq.$\,$(3.13)$]$.
}
\label{fig:2}
\end{figure}

\begin{figure}
\caption{A graphical representation of the two-dimensional Ising
system, where the horizontal and vertical directions correspond, respectively,
the real-space and Trotter directions.  The open circles denote the Ising
variables $\{s_{4\ell-3}^{(r')}\}$ and $\{s_{4\ell-2}^{(r')}\}$, which take
the two values $\pm\frac{1}{2}$, and the open squares denote the Ising
variables $\{S_{4\ell-1}^{(r')}\}$, and $\{S_{4\ell}^{(r')}\}$, which take the
three values $0$ and $\pm 1$. The plaquettes shaded by the slashed,
back-slashed, vertical, and horizontal lines correspond, respectively, to the
local Boltzmann factors,
$\rho_{4\ell-3,4\ell-2}^{(2r-1,2r)}$, $\rho_{4\ell-1,4\ell}^{(2r-1,2r)}$,
$\rho_{4\ell-2,4\ell-1}^{(2r,2r+1)}$, and $\rho_{4\ell,4\ell+1}^{(2r,2r+1)}$.
}
\label{fig:3}
\end{figure}

\begin{figure}
\caption{Plot of the Trotter-extrapolated energy gap $\Delta(N)$ versus $1/N$
with $N\!=\!16$, $32$, $64$, and $128$, for (a) $J_2\!=\!1.00$ (the open
circles), $J_2\!=\!0.90$ (the open squares), $J_2\!=\!0.80$ (the open
diamonds), and $J_2\!=\!0.78$ (the open triangles) and for (b) $J_2\!=\!0.75$
(the closed triangles), $J_2\!=\!0.70$ (the closed diamonds), $J_2\!=\!0.60$
(the closed squares), and $J_2\!=\!0.50$ (the closed circles), where
$J_1\!=\!J_3\!=\!1.0$.  The solid lines give curves of a least-squares fit to
eq.$\,$(3.12) (see the text for more details).
}
\label{fig:4}
\end{figure}

\begin{figure}
\caption{Plot versus $J_2$ of the $N\!\to\!\infty$ extrapolated ground-state
energy per unit cell $\epsilon_{\rm g}(\infty)$ , where
$J_1\!=\!J_3\!=\!1.0$.  The solid line is a guide to the eye.
}
\label{fig:5}
\end{figure}

\begin{figure}
\caption{Plot versus $J_2$ of the $N\!\to\!\infty$ extrapolated energy gap
$\Delta(\infty)$, where $J_1\!=\!J_3\!=\!1.0$.  The solid lines are guides to
the eye.
}
\label{fig:6}
\end{figure}

\begin{figure}
\caption{Plots versus $J_2$ of the \hbox{$N\!\to\!\infty$} extrapolated
ground-state nearest-neighbor two-spin correlation functions
$\omega_{1,2}(\infty)$ (the open circles), $\omega_{2,3}(\infty)$ (the open
squares), and $\omega_{3,4}(\infty)$ (the open diamonds), each of which is
normalized by its minimum value.  Note that $J_1\!=\!J_3\!=\!1.0$.  The solid
lines are guides to the eye.
}
\label{fig:7}
\end{figure}

\begin{figure}
\caption{Schematic representations of the ground states for (a)
$J_2\!<\!J_{2{\rm c}}$ and (b) $J_2\!>\!J_{2{\rm c}}$ by means of the VBS
picture,~\cite{rf:20} where $J_1\!=\!J_3\!=\!1.0$.  The solid circles
represent the $S\!=\!1/2$ spins, and two $S\!=\!1/2$ spins connected by the
solid line form a singlet pair.  Each open ellipse surrounding two $S\!=\!1/2$
represents an operation of constructing an $S\!=\!1$ spin from these
$S\!=\!1/2$ spins by symmetrizing them.
}
\label{fig:8}
\end{figure}

\begin{figure}
\caption{Plot versus the temperature $T$ of (a) the Trotter-extrapolated
specific heat $C(N)/(N/4)$ per unit cell and (b) the Trotter-extrapolated
magnetic susceptibility $\chi(N)/(N/4)$ per unit cell with $N\!=\!8$ (the open
circles), $16$ (the crosses), and $32$ (the open diamonds) for
$J_2\!=\!1.0$.  The dotted lines in (a) and (b) show, respectively, the exact
results for the specific heat per unit cell and the magnetic susceptibility
per unit cell, for $N\!=\!8$ and $J_2\!=\!1.0$.  Note that
$J_1\!=\!J_3\!=\!1.0$.
}
\label{fig:9}
\end{figure}

\begin{figure}
\caption{Plot versus the temperature $T$ of (a) the Trotter-extrapolated
specific heat $C(N)/(N/4)$ per unit cell and (b) the Trotter-extrapolated
magnetic susceptibility $\chi(N)/(N/4)$ per unit cell with $N\!=\!8$ (the open
circles), $16$ (the crosses), and $32$ (the open diamonds), $64$ (the open
squares), and $128$ (the open triangles) for $J_2\!=\!0.77$, being the
critical value of $J_2$.  Here, $C(N)$ and $\chi(N)$ with $N\!=\!64$ and
$128$ are given only when $k_{\rm B}T\!\leq\!0.50$.  The dotted lines in (a)
and (b) show, respectively, the exact results for the specific heat per unit
cell and the magnetic susceptibility per unit cell, for $N\!=\!8$ and
$J_2\!=\!0.77$.  Note that $J_1\!=\!J_3\!=\!1.0$.
}
\label{fig:10}
\end{figure}

\begin{figure}
\caption{Plot versus the temperature $T$ of (a) the Trotter-extrapolated
specific heat $C(N)/(N/4)$ per unit cell and (b) the Trotter-extrapolated
magnetic susceptibility $\chi(N)/(N/4)$ per unit cell with $N\!=\!8$ (the open
circles), $16$ (the crosses), and $32$ (the open diamonds) for
$J_2\!=\!5.0$.  The dotted lines in (a) and (b) show, respectively, the exact
results for the specific heat per unit cell and the magnetic susceptibility
per unit cell, for $N\!=\!8$ and $J_2\!=\!5.0$.  Note that
$J_1\!=\!J_3\!=\!1.0$.  The dot-dashed lines in (a) and (b) show,
respectively, the specific heat and the magnetic susceptibility for two pairs
of the two-spin system consisting of an $S\!=\!1/2$ spin $\vecs$ and an
$S\!=\!1$ spin $\vecS$ which couple with each other by
$5\vecs\cdot\vecS$.  Furthermore, the dashed line in (a) shows the half of
the specific heat for the $S\!=\!1/2$ bond-alternating chain described by the
Hamiltonian ${\cal H}_{(a)}$ given by eq.$\,$(2.3) with $J_1\!=\!J_3\!=\!1.0$
and $N\!=\!8$.
}
\label{fig:11}
\end{figure}
\end{document}